\documentclass[article,nojss]{jss}\usepackage[]{graphicx}\usepackage[]{color}
\makeatletter
\def\maxwidth{ %
  \ifdim\Gin@nat@width>\linewidth
    \linewidth
  \else
    \Gin@nat@width
  \fi
}
\makeatother

\usepackage{Sweave}

\usepackage{myjss}  

\author{Haziq Jamil\\Universiti Brunei Darussalam
   \And Wicher Bergsma\\London School of Economics}
\Plainauthor{Haziq Jamil, Wicher Bergsma}

\title{\pkg{iprior}: An \proglang{R} Package for Regression Modelling using I-priors}
\Plaintitle{iprior: An R Package for Regression Modelling using I-priors}
\Shorttitle{Regression Modelling using \pkg{iprior}}

\Abstract{
This is an overview of the \proglang{R} package \pkg{iprior}, which implements a unified methodology for fitting parametric and nonparametric regression models, including additive models, multilevel models, and models with one or more functional covariates.
Based on the principle of maximum entropy, an I-prior is an objective Gaussian process prior for the regression function with covariance kernel equal to its Fisher information.
The regression function is estimated by its posterior mean under the I-prior, and hyperparameters are estimated via maximum marginal likelihood.
Estimation of I-prior models is simple and inference straightforward, while small and large sample predictive performances are comparative, and often better, to similar leading state-of-the-art models.
We illustrate the use of the \pkg{iprior} package by analysing a simulated toy data set as well as three real-data examples, in particular, a multilevel data set, a longitudinal data set, and a dataset involving a functional covariate.
}

\Keywords{Gaussian, process, regression, objective, prior, empirical, Bayes, RKHS, kernel, EM, algorithm, Nystr\"om, random, effects, multilevel, longitudinal, functional, I-prior, \proglang{R}}
\Plainkeywords{Gaussian, process, regression, objective, prior, empirical, Bayes, RKHS, kernel, EM, algorithm, Nystrom, random, effects, multilevel, longitudinal, functional, I-prior, R}

\Address{
  Haziq Jamil\\
  Universiti Brunei Darussalam\\
  Mathematical Sciences\\
  Faculty of Science\\
  Jalan Tungku Link\\
  Gadong, BE1410\\
  Brunei Darussalam\\
  E-mail: \email{haziq.jamil@ubd.edu.bn}\\
  URL: \url{https://haziqj.ml/}\\
  \\
  Wicher Bergsma\\
  London School of Economics and Political Science\\
  Department of Statistics\\
  Columbia House\\
  Houghton Street \\
  London WC2A 2AE \\
  United Kingdom \\
  E-mail: \email{w.p.bergsma@lse.ac.uk}\\
  URL: \url{http://stats.lse.ac.uk/bergsma/}
}
\IfFileExists{upquote.sty}{\usepackage{upquote}}{}
\begin{document}


\section{Introduction}

For subject $i \in \{1,\dots, n\}$, assume a real-valued response $y_i$ has been observed, as well as a row vector of $p$ covariates $x_i$ = ($x_{i1},\dots,x_{ip}$), where each $x_{ik}$ belongs to some set $\cX_k$, $k = 1,\dots,p$.
To describe the dependence of the $y_i$ on the $x_i$, we consider the regression model
\begin{align}
  \begin{gathered}\label{eq:linmod}
    y_i = \alpha + f(x_i) + \epsilon_i \\
    \epsilon_i \iid \N(0, \psi^{-1})
  \end{gathered}
\end{align}
where $f$ is some regression function to be estimated, and $\alpha$ is an intercept.
When $f$ can be parameterised linearly as $f(x_i) = x_i^\top \beta$, $\beta \in \bbR^p$, we then have the ordinary linear regression---a staple problem in statistics and other quantitative fields.
We might also have that the data is separated naturally into groups or levels by design, for example, data from stratified sampling, students within schools, or longitudinal measurements over time.
In such a case, we might want to consider a regression function with additive components
\[
  f(x_{ij}, j) = f_1(x_{ij}) + f_2(j) + f_{12}(x_{ij}, j)
\]
where $x_{ij}$ denotes the $i$th observation in group $j\in\{1,\dots,m\}$.
Again, assuming a linear parameterisation, this is recognisable as the multilevel or random-effects linear model, with $f_2$ representing the varying intercept via $f_2(j) = \beta_{0j}$, $f_{12}$ representing the varying slopes via $f_{12}(x_{ij},j) = x_{ij}^\top \beta_{1j}$, and $f_1$ representing the linear component as above.

Moving on from linear models, smoothing models may be of interest as well.
A myriad of models exist for this type of problem, with most classed as nonparametric regression, and the more popular ones are LOcal regrESSion (LOESS), kernel regression, and smoothing splines.
Semiparametric regression models combines the linear component with a non-parameteric component.

Further, the regression problem is made interesting when the set of covariates $\cX$ is functional---in which case the linear regression model aims to estimate coefficient functions $\beta:\cT \to \bbR$ from the model
\[
  y_i = \int_\cT x_i(t)\beta(t)\d t + \epsilon_i.
\]
Nonparametric and semiparametric regression with functional covariates have also been widely explored.

On the software side, there doesn't seem to be a shortage of packages in \proglang{R} \citep{R} to fit the models mentioned above.
The \code{lm()} function provides simple and multiple linear regression in base \proglang{R}.
For multilevel models, \pkg{lme4} \citep{bates2014lme4} is widely used.
These two are likelihood based methods, but Bayesian alternatives are available as well in \pkg{rstanarm} \cite{rstanarm} and \pkg{brms} \citep{buerkner2016brms}.
For smoothing models, base \proglang{R} provides the functions \code{smooth.spline()} for modelling with smoothing splines, and \code{ksmooth()} for Nadaraya-Watson kernel regression.
The \pkg{mgcv} package \citep{mgcv} is an extensive package that is able to fit (generalised) additive models.
Finally, the \pkg{fda} package \citep{fda} fits functional regression models in \proglang{R}.

The \pkg{iprior} package provides a platform in \proglang{R} to estimate a wide-range of regression models, including the ones described above, using what we call the I-prior methodology.
As such, it can be seen as a unifying methodology for various regression models.
Estimation and inference is simple and straightforward using maximum likelihood, and thus the package provides the end-user with the tools necessary to analyse and interpret various types of regression models.
Prediction is also possible, with small and large sample performance comparative to, though often better than, other methodologies such as the closely related Gaussian process regression.

\subsection{The I-prior regression model}

For the regression model stated in \eqref{eq:linmod}, we assume that the function $f$ lies in a reproducing kernel Hilbert space (RKHS) of functions $\cF$ with reproducing kernel $h:\cX \times \cX \to \bbR$.
Often, the reproducing kernel (or simply kernel, for short) is indexed by one or more parameters which we shall denote as $\eta$.
Correspondingly, the kernel is rightfully denoted as $h_\eta$ to indicate the dependence of the parameters on the kernels, though where this is seemingly obvious, might be omitted.

The definition of an RKHS entails that any function in $\cF$ can be approximated arbitrarily well by functions of the form
\begin{align}\label{eq:ipriorre}
  f(x) = \sum_{i=1}^n h(x,x_i)w_i
\end{align}
where $w_1,\dots,w_n$ are real-valued\footnotemark.
We define the \emph{I-prior} for our regression function $f$ in~(\ref{eq:linmod}) as the distribution of a random function of the form \eqref{eq:ipriorre} when each of the $w_i$ are independently distributed as $\N(0,\psi)$, with $\psi$ being the model error precision.
As a result, we may view the I-prior for $f$ as having the Gaussian process distribution
\begin{align}\label{eq:iprior}
  \bff = \big(f(x_1),\dots,f(x_n) \big)^\top \sim \N_n(\bzero, \psi\bH_\eta^2)
\end{align}
with $\bH_{\eta}$ an $n \times n$ matrix with $(i,j)$ entries equal to $h_\eta(x_i,x_j)$, and $\bzero$ a length $n$ vector of zeroes.
The covariance kernel of this prior distribution is related to the Fisher information for $f$ \citep{bergsma2017}, and hence the name I-prior---the `I' stands for information.
The prior mean of zero for the regression function is typically reasonable, and this is what the \pkg{iprior} package implements.

\footnotetext{That is to say, $\cF$ is spanned by the functions $h(\cdot,x)$. More precisely, $\cF$ is the completion of the space $\cG = \text{span}\{h(\cdot,x) | x \in \cX \}$ endowed with the squared norm $\norm{f}^2_\cG = \sum_{i=1}^n\sum_{i=1}^n w_i w_j h(x_i,x_j)$ for $f$ of the form \eqref{eq:ipriorre}. See, for example, \cite{berlinet2011reproducing} for details.}

As with Gaussian process regression (GPR), the function $f$ is estimated by its posterior mean.
In the normal model, the posterior distribution for the regression function conditional on the responses $\by = (y_1,\dots,y_n)$,
\begin{align}
  p(\bff|\by) = \frac{p(\by|\bff)p(\bff)}{\int p(\by|\bff)p(\bff) \d \bff},
\end{align}
can easily be found, and it is in fact normally distributed.
The posterior mean for $f$ evaluated at a point $x \in \cX$ is given by
\begin{align}\label{eq:postmean}
  \E\big[f(x)|\by\big] = \bh_\eta^\top(x) \cdot
  {\color{gray}
  \overbrace{\color{black} \psi\bH_\eta\big(\psi\bH_\eta^2 + \psi^{-1}\bI_n\big)^{-1}\by}^{\tilde \bw}
  }
\end{align}
where we have defined $\bh_\eta^\top(x)$ to be the vector of length $n$ with entries $h_\eta(x,x_i)$ for $i=1,\dots,n$.
Incidentally, the elements of the $n$-vector $\tilde \bw$ defined in \eqref{eq:postmean} are the posterior means of the random variables $w_i$ in the formulation \eqref{eq:ipriorre}.
The posterior variance of $f$ is given by
\begin{align}
  \VAR\big[f(x)|\by\big] = \bh_\eta^\top(x)\big(\psi\bH_\eta^2 + \psi^{-1}\bI_n\big)^{-1}\bh_\eta^\top(x).
\end{align}
These are of course well-known results in Gaussian process literature---see, for example, \cite{rasmussen2006gaussian} for details.


\subsection{Estimation}

The kernel parameter $\eta$ and the error precision $\psi$ (which we collectively refer to as the model hyperparameters of the covariance kernel $\theta$) can be estimated in several ways.
One of these is direct optimisation of the marginal log-likelihood---the most common method in the Gaussian process literature.
\begin{align*}
  \log L(\theta)
  &= \log \int p(\by|\bff)p(\bff) \d \bff \\
  &= -\half[n]\log 2\pi - \half\log\vert \bSigma_\theta \vert - \half \by^\top \bSigma_\theta^{-1} \by
\end{align*}
where $\bSigma_\theta = \psi\bH_\eta^2 + \psi^{-1}\bI_n$.
This is typically done using conjugate gradients with a Cholesky decomposition on the covariance kernel to maintain stability, but the \pkg{iprior} package opts for an eigendecomposition of the kernel matrix (Gram matrix) $\bH_{\eta} = \bV\cdot\text{diag}(u_1,\dots,u_n)\cdot\bV^\top$ instead.
Since $\bH_{\eta}$ is a symmetrix matrix, we have that $\bV\bV^\top = \bI_n$, and thus
\[
  \bSigma_\theta = \bV \cdot \text{diag} (\psi u_1^2 + \psi^{-1},\dots,\psi u_n^2 + \psi^{-1}) \cdot \bV^\top
\]
for which the inverse and log-determinant is easily obtainable.
This method is relatively robust to numerical instabilities and is better at ensuring positive definiteness of the covariance kernel.
The eigendecomposition is performed using the \pkg{Eigen} \proglang{C++} template library and linked to \pkg{iprior} using \pkg{Rcpp} \citep{eddelbuettel2011rcpp}.
The hyperparameters are transformed by the \pkg{iprior} package so that an unrestricted optimisation using the quasi-Newton L-BFGS algorithm provided by \code{optim()} in \proglang R.
Note that minimisation is done on the deviance scale, i.e., minus twice the log-likelihood.
The direct optimisation method can be prone to local optima, in which case repeating the optimisation at different starting points and choosing the one which yields the highest likelihood is one way around this.

Alternatively, the expectation-maximisation (EM) algorithm may be used to estimate the hyperparameters, in which case the I-prior formulation in \eqref{eq:ipriorre} is convenient.
Substituting this into \eqref{eq:linmod} we get something that resembles a random effects model.
By treating the $w_i$ as ``missing'', the $t$th iteration of the E-step entails computing
\begin{align}
  Q(\theta) = \E \left[ \log p(\by, \bw | \theta) \big\vert \by,\theta^{(t)} \right].
\end{align}
As a consequence of the properties of the normal distribution, the required joint and posterior distributions $p(\by, \bw)$ and $p(\bw | \by)$ are easily obtained.
The M-step then maximises the $Q$ function above, which boils down to solving the first order conditions
\begin{align}
  \frac{\partial Q}{\partial\eta}
  &= -\half \tr \left(\frac{\partial \bSigma_\theta}{\partial\eta} \tilde\bW^{(t)} \right) + \psi \cdot \by ^\top \frac{\partial \bH_\eta}{\partial\eta} \tilde\bw^{(t)} \label{eq:emtheta} \\
  \frac{\partial Q}{\partial\psi}
  &= -\half \by^\top\by - \tr \left(\frac{\partial \bSigma_\theta}{\partial\psi} \tilde\bW^{(t)} \right) + \by^\top \bH_\eta \tilde\bw^{(t)} \label{eq:empsi}
\end{align}
equated to zero.
Here, $\tilde\bw$ and $\tilde\bW$ are the first and second posterior moments of $\bw$.
The solution to \eqref{eq:empsi} can be found in closed-form, but not necessarily for \eqref{eq:emtheta}.
In cases where closed-form solutions exist, then it is just a matter of iterating the update equations until a suitable convergence criterion is met (e.g. no more sizeable increase in successive log-likelihood values).
In cases where closed-form solutions do not exist for $\theta$, the $Q$ function is again optimised with respect to $\theta$ using the L-BFGS algorithm.

The EM algorithm is more stable than direct maximization, and is especially suitable if there are many scale parameters. However, it is typically slow to converge.
The \pkg{iprior} package provides a method to automatically switch to the direct optimisation method after running several EM iterations.
This then combines the stability of the EM with the speed of direct optimisation.

For completeness, it should be mentioned that a full Bayesian treatment of the model is possible, with additional priors on the hyperparameters set.
Markov chain Monte Carlo (MCMC) methods can then be employed to sample from the posteriors of the hyperparameters, with point estimates obtained using the posterior mean or mode, for instance.
Additionally, the posterior distribution encapsulates the uncertainty about the parameter, for which inference can be made.
Posterior sampling can be done using Gibbs-based methods in \pkg{WinBUGS} \citep{lunn2000winbugs} or \pkg{JAGS} \citep{plummer2003jags}, and both have interfaces to \proglang{R} via \pkg{R2WinBUGS} \citep{sturtz2005r2winbugs} and \pkg{runjags} \citep{denwood2016runjags} respectively.
Hamiltonian Monte Carlo (HMC) sampling is also a possibility, and the \proglang{Stan} project \citep{carpenter2016stan} together with the package \pkg{rstan} \citep{rstan}  makes this possible in \proglang{R}.
All of these MCMC packages require the user to code the model individually, and we are not aware of the existence of MCMC-based packages which are able to estimate GPR models.
This makes it inconvenient for GPR and I-prior models, because in addition to the model itself, the kernel functions need to be coded as well and ensuring computational efficiency would be a difficult task.
Note that this full Bayesian method is not implemented in \pkg{iprior}, but described here for completeness.

Finally, a note on the intercept.
Given the regression model \eqref{eq:linmod} subject to an I-prior \eqref{eq:iprior}, the marginal likelihood of the intercept $\alpha$ (after integrating out the I-prior) can be maximised with respect to $\alpha$, which yields the sample mean for $y$ as the ML estimate for intercept.

\subsection{Computational considerations}
\label{sec:computational}

Computational complexity for estimating I-prior models (and in fact, for GPR in general) is dominated by the inversion of the $n \times n$ matrix $\bSigma_\theta = \psi\bH_\eta^2 + \psi^{-1}\bI_n$, which scales as $O(n^3)$ in time.
As mentioned earlier, the \pkg{iprior} package inverts this by way of the eigendecomposition of $\bH_\eta$, but this operation is also $O(n^3)$.
For the direct optimisation method, this matrix inversion is called when computing the log-likelihood, and thus must be computed at each Newton step.
For the EM algorithm, this matrix inversion appears when calculating $\tilde \bw$, the posterior mean of the I-prior random effects.
Furthermore, storage requirements for I-priors models are similar to that of GPR models, which is $O(n^2)$.

The machine learning literature is rich in ways to resolve this issue, as summarised by \cite{quinonero2005unifying}.
One such method is to use low-rank matrix approximations.
Let $\bQ$ be a matrix with rank $q < n$, and that $\bQ\bQ^\top$ can be used to approximate the kernel matrix $\bH_\eta$.
Then
\[
  (\psi\bH_\eta^2 + \psi^{-1}\bI_n)^{-1} \approx
  \psi\left[
  \bI_n -
  \bQ\left( \big(\psi^2\bQ^\top\bQ\big)^{-1} +\bQ^\top\bQ \right)^{-1} \bQ^\top
  \right],
\]
obtained via the Woodbury matrix identity, is a potentially much cheaper operation which scales $O(nq^2)$---$O(q^3)$ to do the inversion, and $O(nq)$ to do the multiplication (because typically the inverse is premultiplied to a vector).
When the kernel matrix itself is sufficiently low ranked (for instance, when using the linear kernel for a low-dimensional covariate) then the above method is exact.
However, other interesting kernels such as the fractional Brownian motion (fBm) kernel or the squared exponential kernel results in kernel matrices which are full rank.

Another method of approximating the kernel matrix, and the method implemented by our package, is the Nystr\"om method \citep{williams2001using}.
The theory has its roots in approximating eigenfunctions, but this has since been adopted to speed up kernel machines.
The main idea is to obtain an (approximation to the true) eigendecomposition of $\bH_\eta$ based on a small subset $m \ll n$ of the data points.
Reorder the rows and columns and partition the kernel matrix as
\[
  \bH_\eta =
  \begin{pmatrix}
    \bA_{m\times m}         & \bB_{m \times (n-m)} \\
    \bB_{m \times (n-m)}^\top  & \bC_{(n-m) \times (n-m)} \\
  \end{pmatrix}.
\]
The Nystr\"om method provides an approximation to the lower right block $\bC$ by manipulating the eigenvectors and eigenvalues of $\bA$, an $m \times m$ matrix, together with the matrix $\bB$ to give
\[
  \bH_\eta \approx
  \begin{pmatrix}
    \bV_m \\
    \bB^\top\bV_m\bU_m^{-1}
  \end{pmatrix}
  \bU_m
  \begin{pmatrix}
    \bV_m^\top & \bU_m^{-1}\bV_m^\top\bB
  \end{pmatrix}
\]
where $\bU_m$ is the diagonal matrix containing the $m$ eigenvalues of $\bA$, and $\bV_m$ is the corresponding matrix of eigenvectors.
An orthogonal version of this approximation is of interest, which has been studied by \cite{fowlkes2001efficient}, which allows us to easily calculate the inverse of $\bSigma_\theta$.
Estimating I-prior models using the Nystr\"om method takes $O(nm^2)$ times and $O(nm)$ storage.

\subsection{Comparison to other software packages}

It is possible to reformulate the I-prior model as a standard GPR model, by using ``squared'' kernels and a suitable reparameterisation of the hyperparameters (for details see Appendix \ref{apx:ipriorgpr}).
With this in mind, there are a number of excellent GPR software packages available which may potentially be used to fit I-prior models.
In this short review, we specifically describe \proglang{R} packages, and briefly outline the limitations for I-prior modelling.

The package \pkg{kernlab} \citep{zeileis2004kernlab} provides a comprehensive toolkit for not just GPR, but also other kernel-based machine learning models.
The package provides what is termed ``dot product primitives (kernels)'', i.e., the functions to calculate kernels and kernel matrices are exposed to the end-user.
Furthermore, the package allows user-defined kernel functions to extend their \proglang{S4} methods which then means I-prior modelling is possible.
However, kernel hyperparameters must be fixed, and therefore it is not possible to estimate the hyperparameter values from the data.

The package \pkg{gptk} \citep{gptk} on the other hand, originally implemented in \proglang{MATLAB}, is able to estimate hyperparameters by log-likelihood maximisation.
Kernel support is minimal, with the main kernel used being the SE kernel.
The authors of the package have also developed another GPR package called \pkg{gprege} \citep{kalaitzis2011simple}, but it seems it is geared towards the specific usage in time-series genetic expression analysis.

The package \pkg{GPfit} \citep{macdonald2015gpfit} focuses effort on developing a clustering based, gradient type, optimisation algorithm to estimate GPR models using the squared exponential kernel.
There is also no option to change or modify this kernel nor to supply a user-defined kernel.

Finally, \pkg{GPFDA} \citep{gpfda} is a specialised package which performs GPR with functional covariates, though unidimensional covariates are also supported.
Kernel support is limited, with only the linear and SE kernel built-in as of the time of writing, though the package does estimate the hyperparameters and the error precision via maximum likelihood (using conjugate gradient methods).

A common theme among all these packages, besides using the SE kernel as standard, is the lack of a feature to estimate kernel parameters.
Though when this feature is available, it is not possible to supply a user-defined kernel in which to perform I-prior modelling.
Apart from \pkg{kernlab}, which only offers the functionality to use fixed hyperparameter values, there does not seem to be a suitable package to estimate I-prior models.



\section[The iprior package]{The \pkg{iprior} package}

There are three main functions in the \pkg{iprior} package.
The first are the various \code{kern_*()} functions to create kernel matrices.
The second is the \code{kernL()} function to ``load the kernels'' and prepare an I-prior model.
The third is the \code{iprior()} function itself, which either takes a prepared kernel-loaded object from \code{kernL()}, or takes in a model directive straight from the user, and proceeds to fit an I-prior model.
A schematic of the package is shown in the figure below.

\begin{figure}[h]
  \centering
  \includegraphics[width=0.95\linewidth]{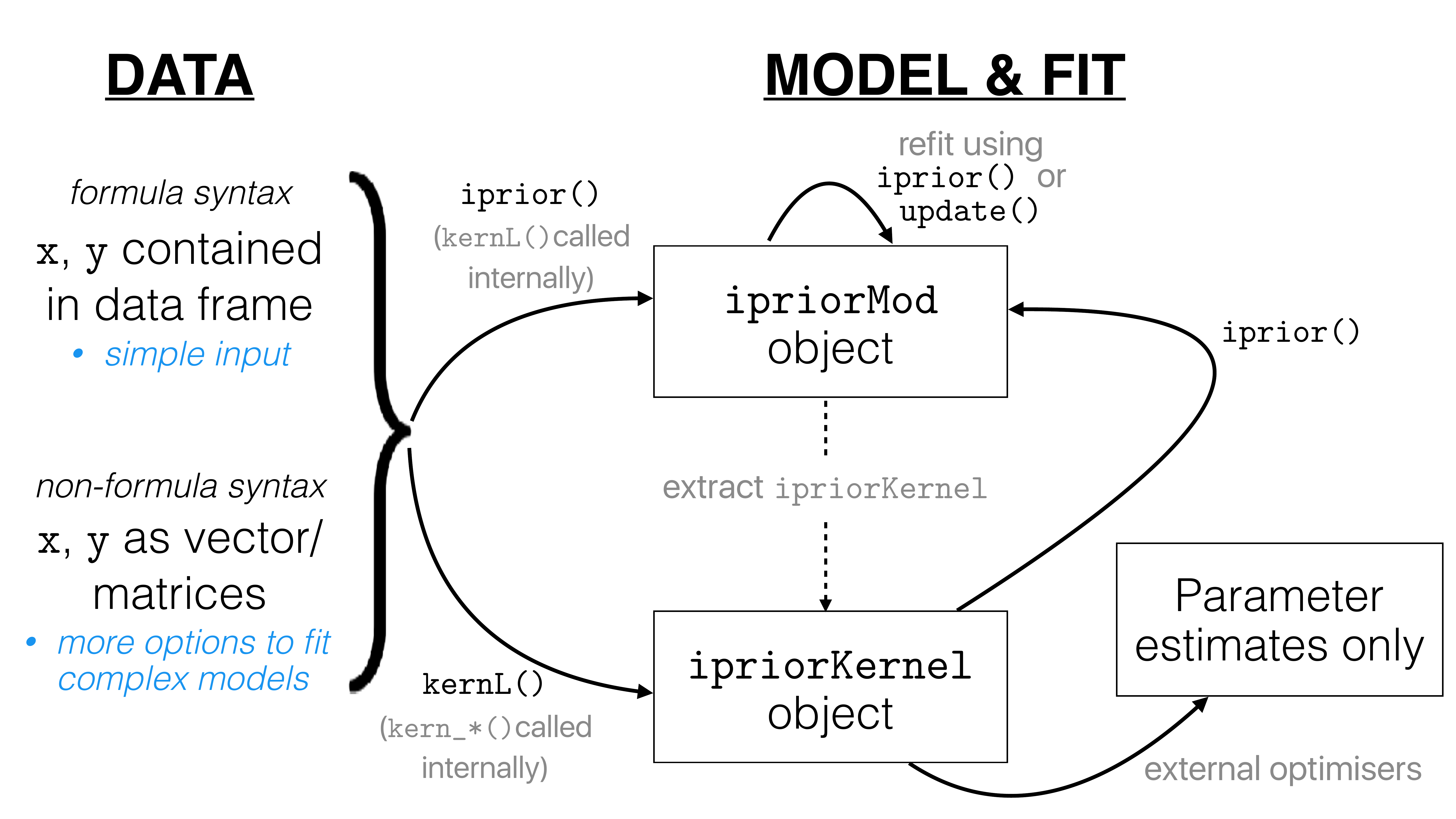}
  \label{fig:packageschematic}
  \caption{A schematic of the package. The main \code{iprior()} function can be called directly, which in turn calls \code{kernL()} internally and that calls the particular \code{kern\_*()} functions. Alternatively, these can be called individually.}
\end{figure}

\subsection{Kernels}

The building blocks of the I-prior models are kernel functions: Symmetric, positive-definite functions which maps pairs of inputs from $\cX$ to a real number, i.e., $h:\cX \times \cX \to \bbR$.
The usefulness of kernels are particularly well-known in the machine learning literature, with many methods taking advantage of what is known as the ``kernel trick'' \citep[Section 6]{bishop2006pattern}.
For our purposes, kernels give rise to reproducing kernel Hilbert spaces (RKHS), which provides us the mathematical structure necessary to perform I-prior modelling.
The choice of kernel determines the space of functions that the I-prior functions resides in, and this package supports five types of RKHSs, which are explained below.
With the exception of the Pearson kernel, the set $\cX$ for the kernels are assumed to be $\bbR^d$, for some $d \in \bbN$.

In what follows, each of the kernel functions have an associated scale parameter $\lambda$.
The scale of the RKHS over the set of covariates may be arbitrary, so scale parameters, typically estimated from the data, are introduced to resolve this.
Setting scale parameter values are optional when calling the kernel functions in \pkg{iprior}, with default values of one.
Some of the kernels are also defined by additional parameters.

\subsubsection{The canonical linear kernel}

The function \code{kern_linear()} (or the alternatively named \code{kern_canonical()})  implements the kernel given by
\[
  h_\lambda(x,x') = \lambda \cdot \langle x,x' \rangle_\cX.
\]
This gives rise to an RKHS of linear functions.

\subsubsection{The fractional Brownian motion (fBm) kernel}

The function \code{kern_fbm()} implements the fractional Brownian motion kernel with Hurst coefficient $\gamma \in (0,1)$, as defined by
\[
  h_{\lambda,\gamma}(x,x') = -\half[\lambda] \left( \Vert x - x' \Vert^{2\gamma}_\cX - \Vert x \Vert^{2\gamma}_\cX - \Vert x' \Vert^{2\gamma}_\cX \right).
\]
This gives rise to functions suitable for smoothing models. The value of the Hurst coefficient acts as a smoothing parameter. The default value $\gamma = 0.5$ is used in the package.

\subsubsection{The squared exponential (SE) kernel}

The function \code{kern_se()} implements the de-facto kernel used in GPR, as defined by
\[
  h_{\lambda,l}(x,x') = \lambda \cdot \exp \left( -\frac{\Vert x - x' \Vert^2_\cX }{2l^2} \right).
\]
The length scale parameter $l>0$ determines how much the smooth exponential functions wiggles.
Parameterised differently, this kernel is also known as the Gaussian kernel or the radial basis function (RBF) kernel.
The default length scale is set to one.

\subsubsection{The $d$-degree polynomial kernel}

The function \code{kern_poly()} implements polynomial kernel with degree $d$ and offset $c \geq 0$, as given by
\[
  h_{\lambda,c,d}(x,x') =  \left( \lambda \cdot \langle x,x' \rangle_\cX + c \right)^d.
\]
This allows modelling with functions with effect similar to that achieved by polynomial regression.
In other words, squared, cubic, or higher order terms may be added to the regression function simply by choosing the correct degree $d$.
The offset parameter may be estimated or left at the default of zero.

\subsubsection{The Pearson kernel}

The so-called Pearson RKHS contains functions that map a countably finite set $\cX$ to the reals, and so the kernel is used for categorical covariates.
Let $\Prob$ be a probability distribution over $\cX$.
The function \code{kern_pearson()} implements the kernel defined by
\[
  h_{\lambda}(x,x') =  \lambda \cdot \left( \frac{\delta_{xx'}}{\Prob(X = x)} - 1 \right),
\]
where $\delta$ is the Kronecker delta.
The package uses the empirical distribution in lieu of the true distribution $\Prob$.
The Pearson RKHS is so named due to its relation with the Pearson chi-square statistic  as seen from the \emph{Hilbert-Schmidt independence criterion} (HSIC) \citep{gretton2005measuring}.

\subsubsection{Examples}

The \code{kern_*()} functions take in vectors or matrices \code{x}, and another optional vector or matrix \code{y}.
If \code{y} is not supplied, then \code{y} is taken to be \code{x}.
The function returns a matrix with entries \code{[i, j]} entries equal to $h(\texttt{x[i, ]}, \texttt{y[j, ]})$.

\begin{Schunk}
\begin{Sinput}
R> # The linear kernel
R> x <- rnorm(5)
R> kern_linear(x, 1:5)
R> # The fBm kernel with Hurst = 0.5
R> y <- rnorm(3)
R> kern_fbm(x, y, gamma = 0.5)
R> # The SE kernel with length scale = 1
R> x <- matrix(rnorm(5 * 3), nrow = 5, ncol = 3)
R> kern_se(x, l = 1)
R> # The polynomial kernel with degree = 2 and offset = 0
R> y <- matrix(rnorm(3 * 3), nrow = 3, ncol = 3)
R> kern_poly(x, y, d = 2, c = 0)
R> # The Pearson kernel
R> x <- factor(1:5)
R> kern_pearson(x)
\end{Sinput}
\end{Schunk}

\begin{Schunk}
\begin{figure}

{\centering \includegraphics[width=7.7cm,height=4.8cm]{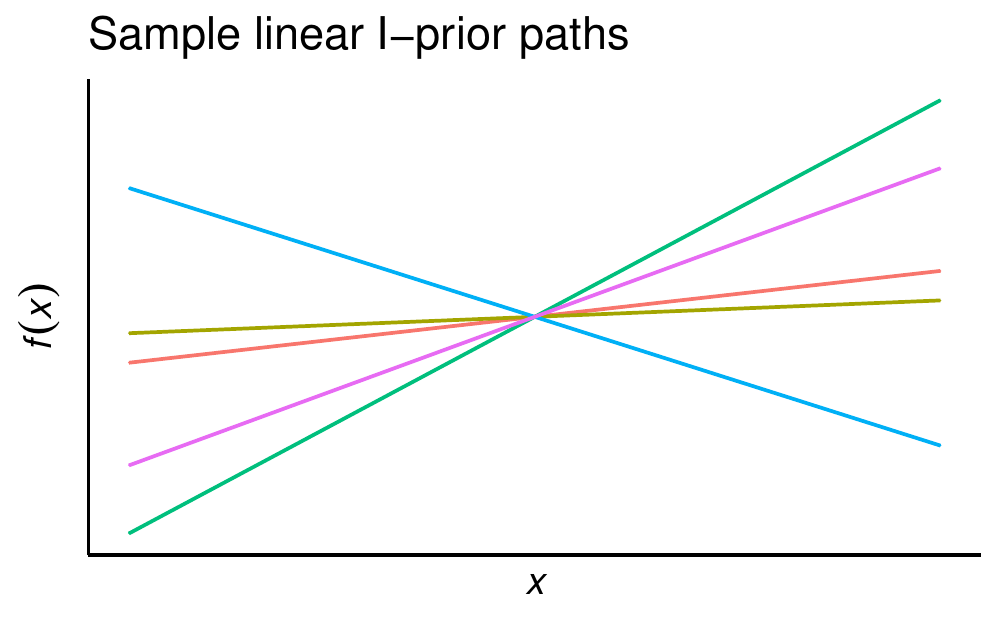} \includegraphics[width=7.7cm,height=4.8cm]{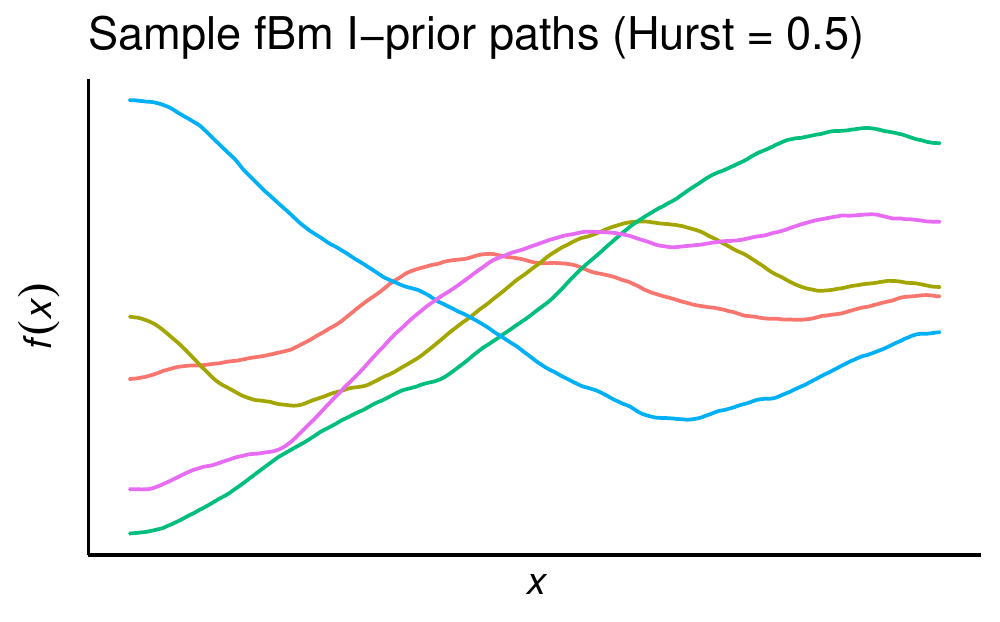} \includegraphics[width=7.7cm,height=4.8cm]{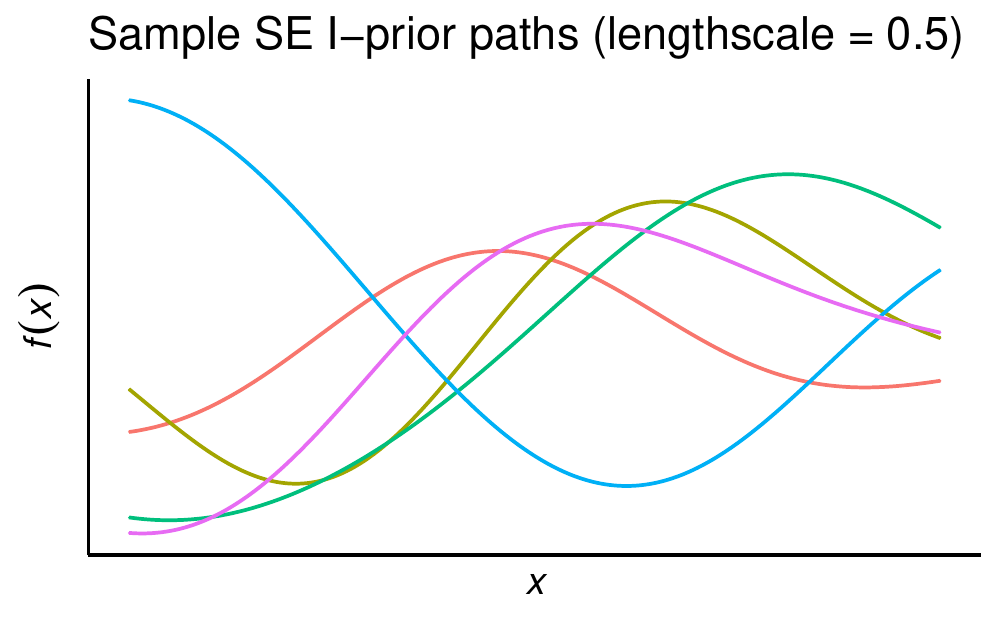} \includegraphics[width=7.7cm,height=4.8cm]{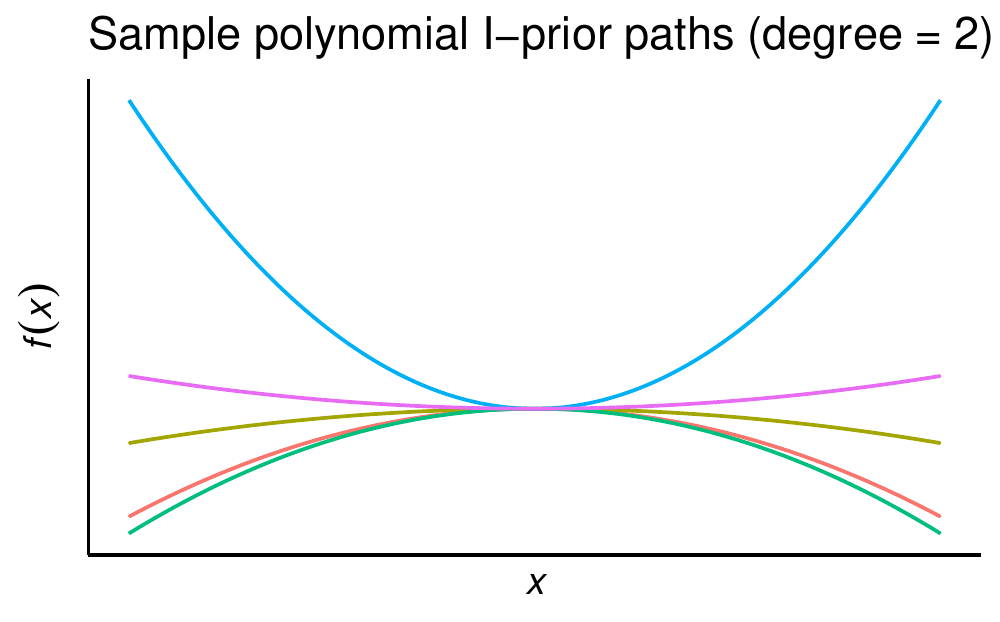} 

}

\caption{Samples of function paths following an I-prior under various kernels. These were generated according to \eqref{eq:ipriorre} with $\psi = 1$.}\label{fig:ipriorpaths}
\end{figure}
\end{Schunk}

\subsubsection{The Sobolev-Hilbert inner product for functional covariates}
\label{sec:sobolevhilbert}

Suppose that we have functional covariates $x$ in the real domain, and that $\cX$ is a set of differentiable functions.
If so, it is reasonable to assume that $\cX$ is a Sobolev-Hilbert space with inner product
\[
  \langle x,x' \rangle_\cX = \int \dot{x}(t) \dot{x}'(t) \d t,
\]
so that we may apply the linear, fBm or any other kernels which make use of inner products by making use of the polarisation identity.
Furthermore, let $z \in \bbR^T$ be the discretised realisation of the function $x \in \cX$ at regular intervals $t = 1,\dots,T$. Then
\[
  \langle x,x' \rangle_\cX \approx \sum_{t=1}^{T-1} (z_{t+1} - z_t)(z'_{t+1} - z_t').
\]
For discretised observations at non-regular intervals then a more general formula to the above one might be used.

\subsubsection{Centred kernels}

As a remark, the package implements \emph{centred versions} of the above kernels when estimating I-prior models, which resolves the possibly arbitrary origin of the RKHS over the set of covariates.
Generally, the centred kernel is defined as
\[
h_{\text{cen}}(x,x') = h(x,x') - \frac{1}{n}\sum_{i=1}^n h(x,x_i) - \frac{1}{n}\sum_{i=1}^n h(x_i,x') + \frac{1}{n^2}\sum_{i=1}^n\sum_{j=1}^n h(x_i,x_j),
\]
such that the sum over columns in the kernel matrix is zero\footnote{The polynomial kernel is not centred this way, but the inner product within it is. The intention of using centred kernels is to achieve centering of the feature space embedding of the data.}.  
The use of centred kernels in the \code{iprior()} function is non-negotiable.
However, when calling the \code{kern_*()} functions independently, an additional option \code{centre = FALSE} may be set to retrieve the non-centred versions of the respective kernels.

\subsection{The kernel loader}

For the remainder of Section 2, we shall be looking at the \code{Orange} data set available in base \proglang{R}, which contains $n = 35$ growth records of the circumference of the trunks of seven orange trees.
The data set also contains the age of the trees (in days) at the time of measurement.
The tree labels are treated as nominal variables, and the rest of the data set as real, continuous measurements.
For simplicity, we shorten the variable names
\begin{Schunk}
\begin{Sinput}
R> names(Orange) <- c("tree", "age", "circ")
R> head(Orange)
\end{Sinput}
\begin{Soutput}
  tree  age circ
1    1  118   30
2    1  484   58
3    1  664   87
4    1 1004  115
5    1 1231  120
6    1 1372  142
\end{Soutput}
\end{Schunk}

\subsubsection{Basic syntax}

The \code{kernL()} function readies an I-prior model for estimation according to the user's specification of the model.
It determines the hyperparameters to be estimated (or fixed), performs the necessary kernel matrix calculations, and outputs an object of class \code{ipriorKernel}.
This can then be passed to the \code{iprior()} function for estimation, which is explained in Section \ref{sec:modelfitting}.
An I-prior model may be specified using formula or non-formula syntax, and the most basic syntax is as follows:
\begin{Schunk}
\begin{Sinput}
R> mod <- kernL(circ ~ age + tree, data = Orange)   # formula syntax
R> with(Orange, mod <<- kernL(y = circ, age, tree))  # non-formula syntax
\end{Sinput}
\end{Schunk}
This would fit the following model:
\begin{align*}
  \begin{gathered}
    \texttt{circ} = \alpha + f(\texttt{age},\texttt{tree}) + \epsilon \\
    \epsilon \sim \N(0,\psi^{-1})
  \end{gathered}
\end{align*}
with $f \in \cF$ (an RKHS), and an I-prior on the regression function $f$.
In the I-prior methodology, we assume an additive decomposition of the regression function into constituent functions dictated by the desired effect of the covariates.
For instance, we could assume that
\[
  f(\texttt{age},\texttt{tree}) = f_1(\texttt{age}) + f_2(\texttt{tree})
\]
where $f_1$ lies in the linear RKHS $\cF_1$ and $f_2$ lies in the Pearson RKHS $\cF_2$, so that $\cF = \cF_1 \oplus \cF_2$.
This would have the effect of regressing separate ``straight line'' functions with similar slopes on the covariate \code{age} for each \code{tree}---in other words, it is a varying-intercept model.
Omitting the \code{tree} variable in the syntax above would fit a linear regression model of \code{circ} on \code{age}.

\subsubsection{Interactions}

A varying-slope effect can be achieved by assuming
\[
  f(\texttt{age},\texttt{tree}) = f_1(\texttt{age}) + f_2(\texttt{tree}) + f_{12}(\texttt{age},\texttt{tree}).
\]
Here, $f_{12}$ is assumed to lie in the so-called tensor product RKHS $\cF_1 \otimes \cF_2$ with kernel $h_{12} = h_1 h_2$.
To fit this model, the additional option \code{interactions} should be called when using non-formula syntax.
In the formula syntax, then use the regular expression for interactions.
\begin{Schunk}
\begin{Sinput}
R> # formula syntax
R> mod <- kernL(circ ~ age + tree + age:tree, data = Orange)
R> # non-formula syntax
R> with(Orange, {
+    mod <<- kernL(y = circ, age, tree, interactions = "1:2")
+  })
\end{Sinput}
\end{Schunk}

The syntax for specifying interactions in the non-formula syntax is \code{"a:b"} to indicate that the variable in position \code{a} interacts with the variable in position \code{b}.
As seen above, this is automatically dealt with when using formula.
The resulting output from \code{print()} contains information regarding the I-prior model prescribed.
\begin{Schunk}
\begin{Sinput}
R> print(mod)
\end{Sinput}
\begin{Soutput}
Sample size: 35 
No. of covariates: 2 
Object size: 235.3 kB

Kernel matrices:
 1 linear [1:35, 1:35] 646646 352329 207584 -65825 -248365 ... 
 2 pearson [1:35, 1:35] 4 4 4 4 4 4 4 -1 -1 -1 ... 
 3 linear x pearson [1:35, 1:35] 2586583 1409318 830335 -263299 -993461 ... 

Hyperparameters to estimate:
lambda[1], lambda[2], psi

Estimation methods available:
direct, em, mixed, fixed
\end{Soutput}
\end{Schunk}
As the output tells us, the I-prior model ``loaded'' has $n = 144$ samples and one covariate using the linear RKHS.
The hyperparameters to estimate are the scale parameters \code{lambda[1]} and \code{lambda[2]}, and the error precision \code{psi}.
Note that since the kernel for the interaction effect is simply the product of the two kernels, there are no additional hyperparameters as a result.
The estimation methods, selectable during the \code{iprior()} fit, are the direct optimisation method, the EM algorithm, and the mixed (EM plus direct) method.
It is also possible just to obtain the posterior regression estimate based on fixed hyperparameters.
Setting user-defined hyperparameter values for the scale parameters, error precision, and other kernel parameters are explained next.

\subsubsection{Specifying kernels and setting hyperparameters}

If instead we assumed that $\cF_1$ is the fBm RKHS, we can achieve a smooth effect of the covariate \code{age} on the response variable.
This is achieved by specifying the \code{kernel} option:
\begin{Schunk}
\begin{Sinput}
R> mod <- kernL(circ ~ ., data = Orange, kernel = "fbm")
R> mod$kernels
\end{Sinput}
\begin{Soutput}
     tree       age 
"pearson" "fbm,0.5" 
\end{Soutput}
\end{Schunk}
Notice that factor type objects are automatically treated with the Pearson kernel, and this is not able to be overridden unless the data is preprocessed beforehand and converted to a \code{numeric} class object.
If there are multiple variables in the model, it is possible to specify them individually, as follows:
\begin{Schunk}
\begin{Sinput}
R> Orange$tree <- as.numeric(Orange$tree)
R> mod <- kernL(circ ~ age + tree, data = Orange,
+               kernel = c("se,0.5", "poly3,1"))
R> get_kernels(mod)
\end{Sinput}
\begin{Soutput}
      age      tree 
 "se,0.5" "poly3,1" 
\end{Soutput}
\end{Schunk}
In this example, the variable \code{tree} has been converted to a \code{numeric} class, and applied a polynomial kernel of degree three and offset equal to one.
Meanwhile, the kernel for \code{age} has been set to a SE kernel with lengthscale 0.5.
Note the use of the comma (\code{,}) to specify the hyperparameter for the kernel.
The syntax is \code{"<kernel name>,<value>"}, where \code{<kernel name>} can be one of \code{fbm}, \code{se} or \code{poly} (the \code{linear} and \code {pearson} kernels do not have hyperparameters associated with them, except the scale).
Omission of the \code{,<value>} is allowed, in which case the default values for the kernels are set.
To set the degree $d$ of the polynomial kernel, use \code{poly} or \code{poly2} for $d = 2$, \code{poly3} for $d = 3$, and so on.

It is also possible to set the values of the scale parameters and error precision by including the arguments \code{lambda = <value>} and/or \code{psi = <value>}, and this is especially relevant if the user would like these values to be treated as fixed.
Note that setting values for any of the hyperparameters above do not indicate that they should not be estimated; this is explained in the next subsection.

\subsubsection{Selecting the hyperparameters to estimate}

With the current five kernels supported by the package, users may select which of the hyperparameters should be estimated in the I-prior model.
This is specified by calling the logical options listed in the Table \ref{tab:est}.
Here's an example:

\begin{Schunk}
\begin{Sinput}
R> (kernL(circ ~ age + tree, Orange, kernel = "fbm", est.hurst = TRUE))
\end{Sinput}
\begin{Soutput}
Sample size: 35 
No. of covariates: 2 
Object size: 39.1 kB

Kernel matrices:
 1 fbm,0.5 [1:35, 1:35] 529 216 87 -107 -205 ... 
 2 pearson [1:35, 1:35] 4 4 4 4 4 4 4 -1 -1 -1 ... 

Hyperparameters to estimate:
lambda[1], lambda[2], hurst[1], psi

Estimation methods available:
direct, em, mixed, fixed
\end{Soutput}
\end{Schunk}

\begin{table}[t!]
\centering
\begin{tabular}{lp{7cm}l}
\hline
Option & Description & Default \\
\hline
\code{est.lambda}      & Estimate scale parameters? & \code{TRUE} \\
\code{est.hurst}       & Estimate fBm Hurst coefficients? & \code{FALSE} \\
\code{est.lengthscale} & Estimate SE lengthscales? & \code{FALSE} \\
\code{est.offset}      & Estimate polynomial offsets? & \code{FALSE} \\
\code{est.psi}         & Estimate the error precision? & \code{TRUE} \\
\code{fixed.hyp}       & Quick \code{TRUE}/\code{FALSE} toggle for all \code{est.*} & \code{NULL} \\
\hline
\end{tabular}
\caption{The various options for which hyperparameters to estimate.}
\label{tab:est}
\end{table}

By default, the scale parameters and error precision are estimated, while the other hyperparameters are not.
Note that the package does not optimise the degree of the polynomial kernel.
To quick set all \code{est.*} options to \code{TRUE}/\code{FALSE}, use the \code{fixed.hyp} option.
When there are several covariates using the same kernel, it is not possible to choose which of the covariates kernel hyperparameters to fix and which to estimate---the current implementation of the package is to either estimate all of them, or to fix all of them.

\subsubsection{The Nystr\"om method}

The Nystr\"om method of approximating the kernel matrix is supported by the package.
This is set by calling the option \code{nystrom = TRUE}.
This would then use a random sample of 10\% of the total data available to estimate the kernel matrix.
Alternatively, the \code{nystrom} option can be set to be a number equal to the Nystr\"om sample size $m$ as described in Section \ref{sec:computational}. Additionally, the seed for random sampling can be controlled by supplying a value to \code{nys.seed}.

\begin{Schunk}
\begin{Sinput}
R> (kernL(circ ~ age, Orange, kernel = "se", nystrom = 10, nys.seed = 123))
\end{Sinput}
\begin{Soutput}
Sample size: 35 
No. of covariates: 1 
Object size: 19.8 kB

Kernel matrices:
 1 se,1 [1:10, 1:35] 0.812 -0.188 -0.188 0.812 0.812 ... 

Hyperparameters to estimate:
lambda, psi

Estimation methods available:
direct (Nystrom), fixed (Nystrom)
\end{Soutput}
\end{Schunk}

\subsection{Model fitting and post-estimation}
\label{sec:modelfitting}

The main function to estimate I-prior models in the form of \code{ipriorKernel} objects is the \code{iprior()} function.
Based on information from the \code{ipriorKernel}, it dispatches the estimation procedure to a selected subroutine and estimates the hyperparameters of the I-prior model, if any.
The resulting fit is an object of class \code{ipriorMod}.
Users may select from \code{"direct"}, \code{"em"}, \code{"mixed"}, or \code{"fixed"} as an option to supply to \code{method} in the \code{iprior()} function.
For instance,

\begin{Schunk}
\begin{Sinput}
R> mod <- kernL(circ ~ age + tree + age:tree, data = Orange)
R> mod.fit <- iprior(mod, method = "direct")  # default method
R> mod.fit <- iprior(mod, method = "em")      # The EM algorithm
\end{Sinput}
\end{Schunk}

The \code{iprior()} function is also a wrapper function for the \code{kernL()} and estimation procedures.
This means that users may call the \code{iprior()} function directly, with the exact same options that are available to \code{kernL()}, without having to invoke the two step procedure of calling \code{kernL()} and then \code{iprior()}.
The following code fits the above model in one step using the EM algorithm and the fBm kernel.

\begin{Schunk}
\begin{Sinput}
R> mod <- iprior(circ ~ age + tree + age:tree, data = Orange, kernel = "fbm",
+                method = "em")
\end{Sinput}
\end{Schunk}

Exposing the kernel loader function for the end user has two advantages. Firstly, I-prior models can sometimes involve high-dimensional matrix multiplications, and these may take a long time to process.
By ``loading the kernel'', the user is able to have a stored \code{ipriorKernel} object which can then be reused and fitted .
A situation where this would be useful would be when the user would like to restart the EM algorithm from different set of starting values, or change to a different estimation procedure.
Rather than having to go through all the kernel loading process all over again, the user can simply call the saved \code{ipriorKernel} object from memory.
The caveat of storing the loaded kernel matrices is that it may take a large amount of memory since I-prior models have $O(n^2)$ storage requirements---this is the price to pay for the convenience.
In practice on modern computers, however, this is unlikely to be a bottleneck.

Secondly, the \code{kernL()} function allows for flexible model fitting.
The \code{logLik()} and \code{deviance()} \proglang{S3} methods can be used on an \code{ipriorKernel} object to calculate the log-likelihood and deviance respectively at the hyperparameter value \code{theta}.
There is flexibility in the sense that the user can then estimate the hyperparameters of the I-prior model through means other than \code{optim()}'s L-BFGS algorithm or the EM algorithm, which is built in to the \code{iprior()} function.
A wide range of optimisation packages are available in \proglang{R}---see  \url{https://cran.r-project.org/web/views/Optimization.html} for details.

\subsubsection{Control options}

Besides the estimation \code{method} and model options from \code{kernL()}, the other main option is the \code{control} of fitting methods.
The user supplies the \code{iprior()} function a list containing the \code{control} options to modify.
The available \code{control} options are:

\begin{enumerate}
  \item \code{maxit}

  The maximum number of iterations for either the L-BFGS optimisation procedure or the EM algorithm.
  Defaults to 100.

  \item \code{em.maxit}

  When using \code{method = "mixed"}, this controls the number of initial EM steps before switching to the direct optimisation method.
  Defaults to 5.

  \item \code{stop.crit}

  The stopping criterion for either the L-BFGS optimisation procedure or the EM algorithm.
  The algorithm terminates when it is not able to improve the log-likelihood value by \code{stop.crit}.
  Defaults to \code{1e-8}.

  \item \code{theta0}

  The initial values for the hyperparameters.
  By default, these are set to random values, but may be changed by the user.
  Note that the hyperparameters have been transformed so that an unconstrained optimisation can be performed---see Appendix \ref{apx:hyperparam} for details.

  \item \code{restarts}

  The estimation procedure can be restarted multiple times from different initial values by setting \code{restarts = TRUE}.
  This is especially useful when local optima are present.
  The run with the highest log-likelihood value is chosen automatically.
  By default, the number of restarts is equal to the number of available cores on the machine, and each run is parallelised on each core.
  This is achieved using the \proglang{R} packages \pkg{foreach} \citep{foreach} and \pkg{doSNOW} \citep{dosnow}.

  \item \code{no.cores}

  The number of cores to make available to \code{iprior()} for the parallel restarts.
  By default, it detects and sets this to the maximum number of cores available on the machine.

  \item \code{par.maxit}

  The \code{maxit} for each parallel run when using \code{restarts}. By default, this is set to 5 only, with the intent of continuing estimation from the run with the highest likelihood value.

  \item \code{silent}

  A logical option to turn on or off progress feedback from the estimation procedures.

\end{enumerate}

An example for setting \code{control} options is shown below.

\begin{Schunk}
\begin{Sinput}
R> mod <- kernL(circ ~ age + tree + age:tree, data = Orange)
R> # Set a higher number of maximum iterations and more lenient stop.crit
R> mod.fit <- iprior(mod, control = list(maxit = 500, stop.crit = 1e-3))
R> # Set the mixed method to run more EM steps
R> mod.fit <- iprior(mod, control = list(em.maxit = 50))
R> # Start from lambda = c(2, 2) and psi = 0.5
R> mod.fit <- iprior(mod, control = list(theta0 = c(2, 2, log(0.5))))
R> # Perform four restarts on four cores
R> mod.fit <- iprior(mod, control = list(restarts = 4, no.cores = 4))
R> # Completely turn off reporting
R> mod.fit <- iprior(mod, control = list(silent = TRUE))
\end{Sinput}
\end{Schunk}

\subsubsection{Refit and update}

The \code{iprior} function is written as an \proglang{S3} generic which is able to dispatch on \code{ipriorMod} objects as well.
A practical situation for this is when the EM algorithm has not fully converged within the \code{maxit} supplied, then one can simply run

\begin{Schunk}
\begin{Sinput}
R> mod.fit <- iprior(mod.fit)
\end{Sinput}
\end{Schunk}

which then estimates the I-prior model from the previous obtained hyperparameter values.
This is possible because the resulting \code{ipriorMod} object also contains the \code{ipriorKernel} object (from \code{kernL()}).
Even better still, running

\begin{Schunk}
\begin{Sinput}
R> update(mod.fit, iter.update = 100)
\end{Sinput}
\end{Schunk}

updates the \code{ipriorMod} object with another 100 iterations using the same estimation method, and then overwrites \code{mod.fit} in environment without having to explicitly do the assignment.
It is also possible to set a new estimation \code{method} and supply a new \code{control} list.

\subsubsection{Post-estimation}

There are several methods written for \code{ipriorMod} objects.
The \code{print()} and \code{summary()} are ways to inspect the resulting fit and obtain information required for analysis and inference.
This includes information on the I-prior model and fit information such as the model call/formula, the kernels used on the covariates, the estimation method and relating convergence information.
The summary also includes the estimated hyperparameter values, along with their standard errors and $p$-values for an asymptotic $Z$-test of normality, the log-likelihood value and training mean squared error.

\begin{Schunk}
\begin{Sinput}
R> mod.fit <- iprior(circ ~ . ^ 2, data = Orange, method = "em",
+                    control = list(maxit = 5000))
\end{Sinput}
\begin{Soutput}
====================================
Converged after 2581 iterations.
\end{Soutput}
\begin{Sinput}
R> print(summary(mod.fit), wrap = TRUE)  # wrap option to neaten LaTeX output
\end{Sinput}
\begin{Soutput}
Call:
iprior(formula = circ ~ .^2, data = Orange, method = "em",
  control = list(maxit = 5000))

RKHS used:
Pearson (tree)
Linear (age)

Residuals:
    Min.  1st Qu.   Median  3rd Qu.     Max. 
-16.7965  -7.2506  -0.5123   7.3863  18.1857 

Hyperparameters:
          Estimate    S.E.      z P[|Z>z|]    
lambda[1]  -9.9940  3.5640 -2.804    0.005 ** 
lambda[2]  -0.0002  0.0001 -2.372    0.018 *  
psi         0.0110  0.0030  3.644   <2e-16 ***
---
Signif. codes:  0 '***' 0.001 '**' 0.01 '*' 0.05 '.' 0.1 ' ' 1

Closed-form EM algorithm. Iterations: 2581/5000 
Converged to within 1e-08 tolerance. Time taken: 2.238565 secs
Log-likelihood value: -160.6596 
RMSE of prediction: 8.882306 (Training)
\end{Soutput}
\end{Schunk}

Inference on the scale parameters $\lambda$ essentially give us a sense of importance of that covariate on the response, since functions in the RKHS are of the form
\[
  f(x) = \lambda \sum_{i=1}^n h(x,x_i) w_i,
\]
as we saw earlier in \eqref{eq:ipriorre}. Thus, a scale parameter which is not statistically significant implies that the contribution of that function can be assumed to be nil.
Furthermore, model comparisons can be done via log-likelihood ratio tests whose test statistic follows an asymptotic $\chi^2$ distribution with degrees of freedom equal to the difference in the number of parameters estimated.
When the number of parameters are the same, then it is a matter of selecting models with higher likelihood.
Such a method of comparing marginal likelihoods can be seen as Bayesian model selection using empirical Bayes factors.

Fitted values may be obtained using \code{fitted()}, and predicted values at a new set of data points \code{newdata} using \code{predict()}.
There is an option so that both function return the credibility interval for the predictions at the \code{alpha} level of significance.
Although the hyperparameters are estimated using maximum likelihood, the regression function itself possesses a posterior distribution, and this is used for the credibility intervals.
The \code{newdata} must be similar to the original data used to fit the model---for formula syntax, this must be a data frame containing identical column names; for non-formula syntax, the variables must be supplied as a list.

\begin{Schunk}
\begin{Sinput}
R> fitted(mod.fit)
\end{Sinput}
\begin{Soutput}
Training RMSE: 8.882306 

Predicted values:
      1       2       3       4       5       6       7       8 
 35.508  65.139  79.711 107.236 125.614 137.029 154.030  33.899 
      9      10 
 79.481 101.898 
# ... with 25 more values
\end{Soutput}
\begin{Sinput}
R> predict(mod.fit, Orange[1:5, ], intervals = TRUE, alpha = 0.05)
\end{Sinput}
\begin{Soutput}
Test RMSE: 6.726375 

Predicted values:
     2.5%    Mean   97.5%
1  12.578  35.508  58.439
2  44.426  65.139  85.851
3  59.653  79.711  99.769
4  87.499 107.236 126.974
5 105.404 125.614 145.824
\end{Soutput}
\end{Schunk}

Several plot functions have been written for \code{ipriorMod} objects, with the \code{plot()} \proglang{S3} method pointing to the \code{plot_fitted()} function.
This plots the fitted regression line through the data points, with the response variable on the $y$-axis and the covariate on the $x$-axis.
Thus, it is only useful when the data is able to be presented on a two-dimensional plane.
The other plot functions are described in Table \ref{tab:methods}.

\begin{table}[t!]
\centering
\begin{tabular}{lp{10cm}}
\hline
\proglang{R} Function & Description \\
\hline
\Top \emph{Methods} \\ \Top
\code{coef}     & Extracts the estimates of the hyperparameters that have been estimated. \\
\code{sigma}    & Extracts the estimate of the model standard deviation of errors (i.e., square root of the inverse error precision). \\
\code{fitted}   & Returns the fitted value of the responses $\hat y_1, \dots, \hat y_n$. \\
\code{predict}  & Calculates fitted values from a new set of covariates. \\
\code{resid}    & Returns the residuals $\hat\epsilon_1, \dots, \hat\epsilon_n$. \\
\code{logLik}   & Returns the log-likelihood value of the fitted I-prior model at the ML estimates. \\
\code{deviance} & Returns twice the negative log-likelihood value. \\
\\
\Top \emph{Accessor functions} \\ \Top
\code{get_intercept} & Obtains the intercept of the regression function. \\
\code{get_hyp} & Obtains all values of the model hyperparameters (both estimated and fixed values).\\
\code{get_lambda} & Obtains scale parameters used for the RKHS. \\
\code{get_psi} & Obtains the model error precision. \\
\code{get_se} & Obtains the standard errors for the estimated hyperparameters. \\
\code{get_kernels} & Obtains the RKHS used for the regression functions. \\
\code{get_kern_matrix} & Obtains the kernel matrix $\bH_\eta$. \\
\code{get_prederror} & Obtains the (root) mean squared error of prediction. \\
\code{get_estl} & Obtains information on which hyperparameters were fixed and which were estimated. \\
\code{get_method} & Obtains the estimation method used to fit the model. \\
\code{get_convergence} & Obtain the convergence information. \\
\code{get_niter} & Obtain the number of iterations (Newton or EM steps) taken to fit the model. \\
\code{get_time} & Obtains the time taken to run the estimation procedure. \\
\code{get_size} & Obtains the size of the \code{ipriorKernel} object (where most of the large matrices are stored). \\
\\
\Top \emph{Plots} \\ \Top
\code{plot} & This currently points to \code{plot_fitted()} for convenience. \\
\code{plot_fitted} & Plot of fitted regression line. \\
\code{plot_resid} & Plot of residuals against fitted values. \\
\code{plot_iter} & Plot of the log-likelihood values over time/iteration. \\
\code{plot_ppc} & Plot of a posterior predictive check of the observed versus fitted distribution of the responses. \\
\hline
\end{tabular}
\caption{Available methods, accessor functions and plot functions for an object of class \code{ipriorMod}.}
\label{tab:methods}
\end{table}

\begin{Schunk}
\begin{figure}

{\centering \includegraphics[width=12cm]{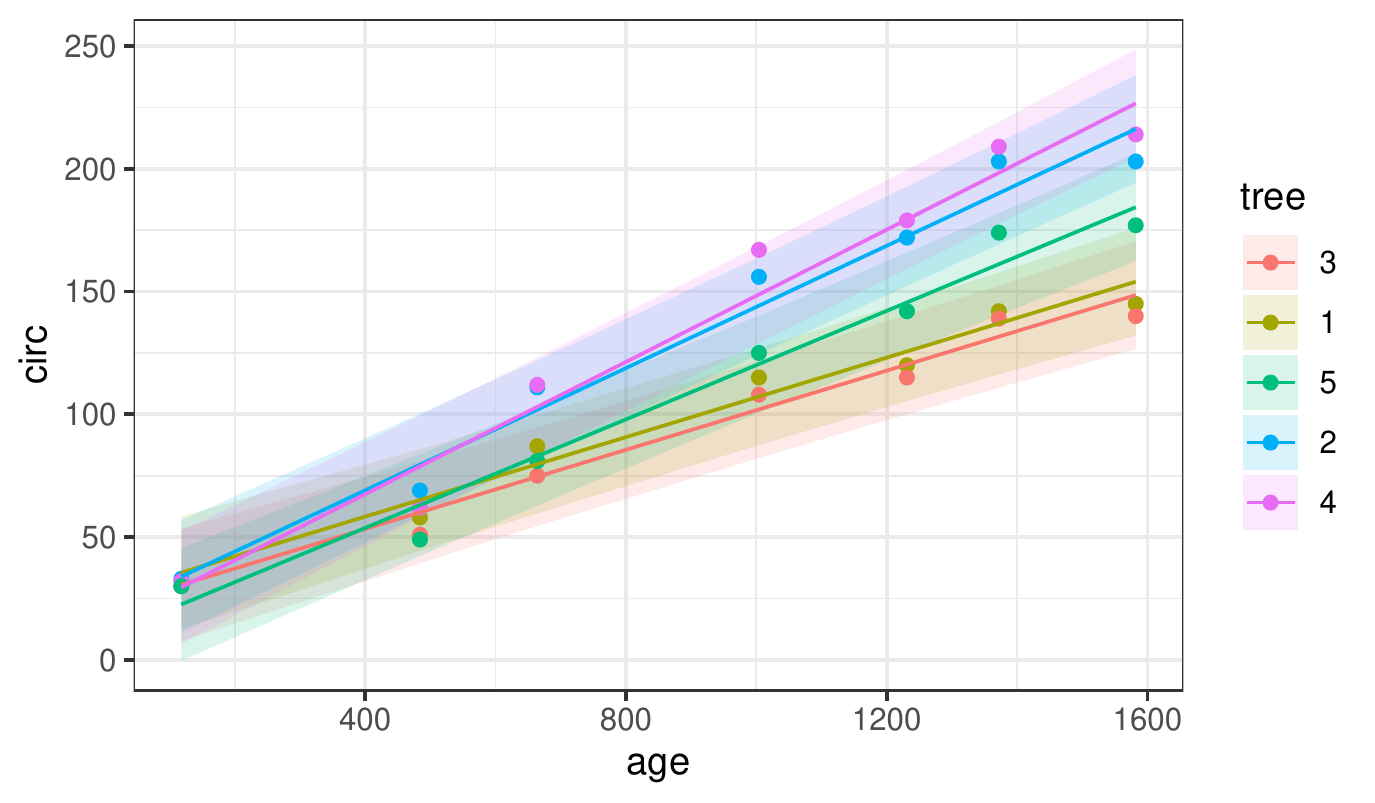} 

}

\caption[Plot of fitted regression line for the \code{Orange} data set]{Plot of fitted regression line for the \code{Orange} data set. Confidence bands for predicted values are also shown.}\label{fig:orangemod}
\end{figure}
\end{Schunk}

\section{Modelling examples}

We demonstrate the use of the \pkg{iprior} package with modelling a toy example from a simulated data set, as well as three other real-data examples.

\subsection[Using the Nystrom method]{Using the Nystr\"om method}

In this section, we investigate the use of the Nystr\"om method of approximating the kernel matrix in estimating I-prior models applied to a toy data set.
The data is obtained by randomly generating data points according to the true regression model
\begin{align*}
  \begin{gathered}
    y_i = \const + \, 0.35 \cdot \phi(x_i;1,0.8^2) + 0.65 \cdot \phi(x_i;4,1.5^2)
    + \ind[x_i > 4.5]\cdot e^{1.25(x_i-4.5)} + \epsilon_i \\
    \epsilon_i \iid \N(0, 0.9 ^ 2)
  \end{gathered}
\end{align*}
where $\phi(x;\mu,\sigma^2)$ represents the PDF of a normal distribution with mean $\mu$ and variance $\sigma^2$.
The features of this regression function are two large bumps at the centres of the mixed Gaussian PDFs, and also a small bump right after $x>4.5$ caused by the additional exponential function.
The true regression function goes to positive infinity as $x$ increases, and to zero as $x$ decreases.
2,000 $(x,y)$ points in the domain $x \in (-1, 5.5)$ have been generated by the built-in \code{gen_smooth()} function, which generates data from the regression model above\footnote{Random fluctuations have also been added to the $(x,y)$ coordinates.}.

\begin{Schunk}
\begin{Sinput}
R> dat <- gen_smooth(n = 2000, xlim = c(-1, 5.5), seed = 1)
R> head(dat)
\end{Sinput}
\begin{Soutput}
            y         X
1  0.05425229 -3.024041
2 -4.42104881 -2.802862
3 -0.80172963 -2.367480
4 -0.26086006 -2.361524
5  1.36148091 -2.319131
6  5.01709840 -2.163912
\end{Soutput}
\end{Schunk}

One could fit the regression model using all available data points, with an I-prior from the fBm-0.5 RKHS of functions as follows (note that the \code{silent} option is used to suppress the output from the \code{iprior()} function):

\begin{Schunk}
\begin{Sinput}
R> (mod.full <- iprior(y ~ X, dat, kernel = "fbm",
+                      control = list(silent = TRUE)))
\end{Sinput}
\begin{Soutput}
Log-likelihood value: -4325.385 

 lambda     psi 
1.35239 0.23638 
\end{Soutput}
\end{Schunk}

To implement the Nystr\"om method, the option \code{nystrom = 50} was added to the above function call, which uses 50 randomly selected data points for the Nystr\"om approximation.

\begin{Schunk}
\begin{Sinput}
R> (mod.nys <- iprior(y ~ X, dat, kernel = "fbm", nystrom = 50,
+                     control = list(silent = TRUE)))
\end{Sinput}
\begin{Soutput}
Log-likelihood value: -1929.554 

 lambda     psi 
0.83283 0.23228 
\end{Soutput}
\end{Schunk}

The hyperparameters estimated for both models are slightly different.
The log-likelihood is also different, but this is attributed to information loss due to the approximation procedure.
Nevertheless, we see from Figure \ref{fig:nystrom.plot} that the estimated regression functions are quite similar in both the full model and the approximated model.
The main difference is that the the Nystr\"om method was not able to extrapolate the right hand side of the plot well, because it turns out that there were no data points used from this region.
This can certainly be improved by using a more intelligent sampling scheme.
The full model took a little under 12 minutes to converge, while the Nystr\"om method took just seconds.
Storage savings is significantly higher with the Nystr\"om method as well.

\begin{Schunk}
\begin{Sinput}
R> get_time(mod.full); get_size(mod.full, units = "MB")
\end{Sinput}
\begin{Soutput}
11.61941 mins
\end{Soutput}
\begin{Soutput}
128.2 MB
\end{Soutput}
\begin{Sinput}
R> get_time(mod.nys); get_size(mod.nys)
\end{Sinput}
\begin{Soutput}
1.438196 secs
\end{Soutput}
\begin{Soutput}
982.2 kB
\end{Soutput}
\end{Schunk}

\begin{Schunk}
\begin{figure}

{\centering \includegraphics[width=7.7cm,height=4.8cm]{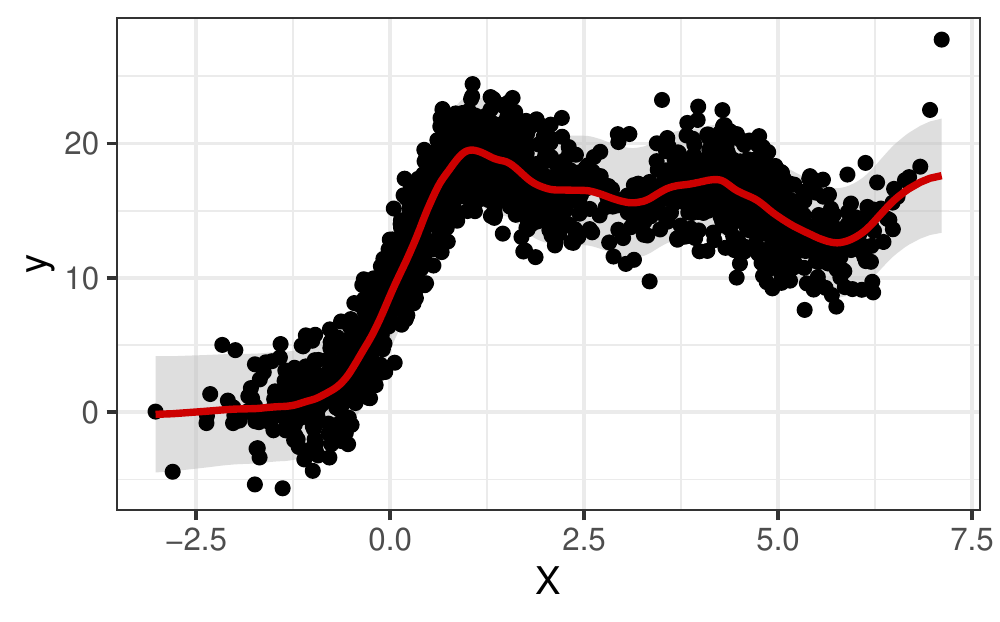} \includegraphics[width=7.7cm,height=4.8cm]{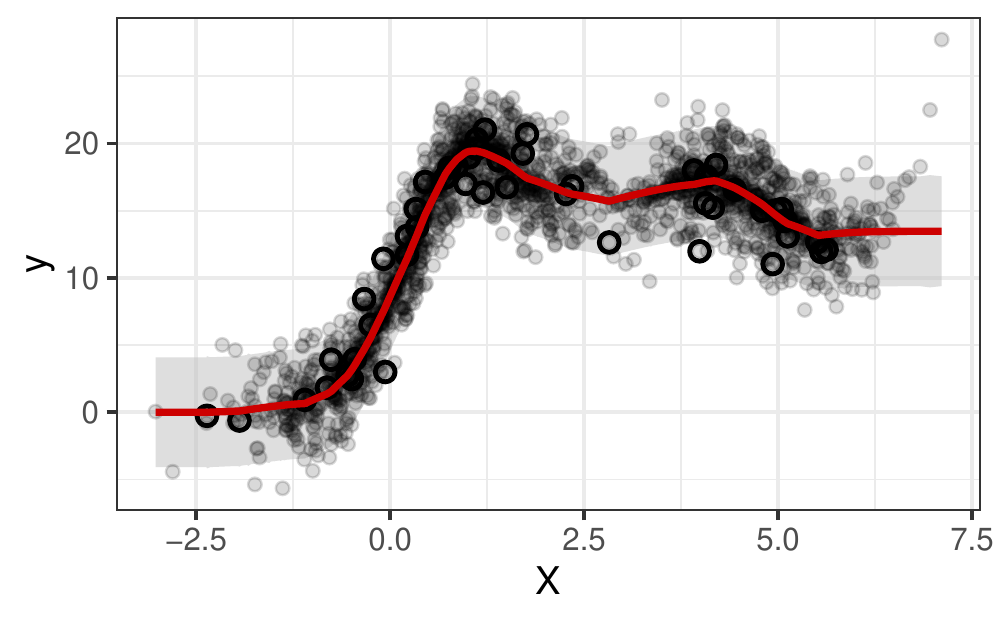} 

}

\caption[Plot of predicted regression function for the full model (left) and the Nystr\"om approximated method (right)]{Plot of predicted regression function for the full model (left) and the Nystr\"om approximated method (right). For the Nystr\"om plot, the data points that were active are shown by circles with bold outlines.}\label{fig:nystrom.plot}
\end{figure}
\end{Schunk}

\subsection{Random effects models}

In this section, a comparison between a standard random effects model and the I-prior approach for estimating varying intercept and slopes model is illustrated.
The example concerns control data\footnotemark\ from several runs of radioimmunoassays (RIA) for the protein insulin-like growth factor (IGF-I) (explained in further detail in \citealt{davidian1995nonlinear}, Section 3.2.1).
RIA is a in vitro assay technique which is used to measure concentration of antigens---in our case, the IGF-I proteins.
When an RIA is run, control samples at known concentrations obtained from a particular lot are included for the purpose of assay quality control.
It is expected that the concentration of the control material remains stable as the machine is used, up to a maximum of about 50 days, at which point  control samples from a new batch is used to avoid degradation in assay performance.

\begin{Schunk}
\begin{Sinput}
R> data(IGF, package = "nlme")
R> head(IGF)
\end{Sinput}
\begin{Soutput}
Grouped Data: conc ~ age | Lot
  Lot age conc
1   1   7 4.90
2   1   7 5.68
3   1   8 5.32
4   1   8 5.50
5   1  13 4.94
6   1  13 5.19
\end{Soutput}
\end{Schunk}

\footnotetext{This data is available in the \proglang{R} package \pkg{nlme} \citep{nlme}.}

The data consists of IGF-I concentrations (\code{conc}) from control samples from 10 different lots measured at differing \code{age}s of the lot.
The data were collected with the aim of identifying possible trends in control values \code{conc} with \code{age}, ultimately investigating whether or not the usage protocol of maximum sample age of 50 days is justified.
\cite{pinheiro2000mixed} remarks that this is not considered a longitudinal problem because different samples were used at each measurement.

We shall  model the IGF data set using the I-prior methodology using the regression function
\[
  f(\texttt{age}, \texttt{Lot}) = f_1(\texttt{age}) + f_2(\texttt{Lot}) + f_{12}(\texttt{age}, \texttt{Lot})
\]
where $f_1$ lies in the linear RKHS $\cF_1$, $f_2$ in the Pearson RKHS $\cF_2$ and $f_{12}$ in the tensor product space $\cF_{12} = \cF_1 \otimes \cF_2$.
The regression function $f$ then lies in the RKHS $\cF = \cF_1 \oplus \cF_2 \oplus \cF_{12}$ with kernel equal to the sum of the kernels from each of the RKHSs\footnote{This is often known as the functional ANOVA decomposition.}.
The explanation here is that the \code{conc} levels are assumed to be related to both \code{age} and \code{Lot}, and in particular, the contribution of \code{age} on \code{conc} varies with each individual \code{Lot}.
This gives the intended effect of a linear mixed-effects model, which is thought to be suitable in this case, in order to account for within-lot and between-lot variability.
We first fit the model using the \pkg{iprior} package, and then compare the results with the standard random effects model using \code{lme4::lmer()}.
The command to fit the I-prior model using the EM algorithm is

\begin{Schunk}
\begin{Sinput}
R> mod.iprior <- iprior(conc ~ age * Lot, IGF, method = "em")
\end{Sinput}
\begin{Soutput}
================================
Converged after 46 iterations.
\end{Soutput}
\begin{Sinput}
R> summary(mod.iprior)
\end{Sinput}
\begin{Soutput}
Call:
iprior(formula = conc ~ age * Lot, data = IGF, method = "em")

RKHS used:
Linear (age)
Pearson (Lot)

Residuals:
   Min. 1st Qu.  Median 3rd Qu.    Max. 
-4.4890 -0.3798 -0.0090  0.2563  4.3972 

Hyperparameters:
          Estimate   S.E.      z P[|Z>z|]    
lambda[1]   0.0000 0.0002 -0.004    0.997    
lambda[2]   0.0007 0.0030  0.239    0.811    
psi         1.4577 0.1366 10.672   <2e-16 ***
---
Signif. codes:  0 '***' 0.001 '**' 0.01 '*' 0.05 '.' 0.1 ' ' 1

Closed-form EM algorithm. Iterations: 46/100 
Converged to within 1e-08 tolerance. Time taken: 2.106749 secs
Log-likelihood value: -291.9033 
RMSE of prediction: 0.8273564 (Training)
\end{Soutput}
\end{Schunk}
\begin{Schunk}
\begin{figure}

{\centering \includegraphics[width=\maxwidth]{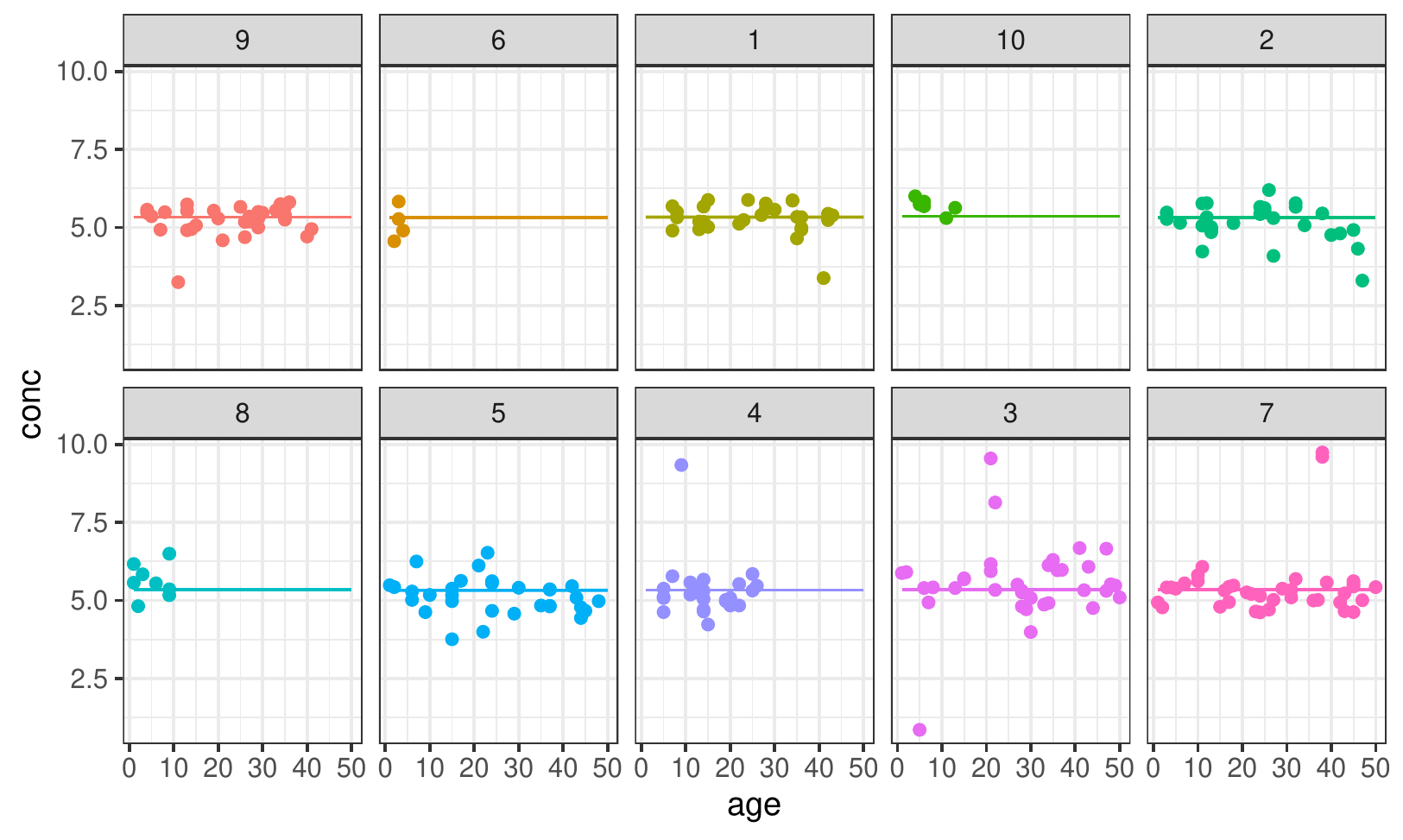} 

}

\caption[Plot of fitted regression line for the I-prior model on the IGF data set, separated into each of the 10 lots]{Plot of fitted regression line for the I-prior model on the IGF data set, separated into each of the 10 lots.}\label{fig:IGF.mod.iprior.plot}
\end{figure}
\end{Schunk}

To make inference on the covariates, we look at the scale parameters \code{lambda}.
We see that both scale parameters for \code{age} and \code{Lot} are close to zero, and a test of significance is not able to reject the hypothesis that these parameters are indeed null.
We conclude that neither \code{age} nor \code{Lot} has a linear effect on the \code{conc} levels.
The plot of the fitted regression line in Figure \ref{fig:IGF.mod.iprior.plot} does show an almost horizontal line for each \code{Lot}.



The standard random effects model, as explored by \cite{davidian1995nonlinear} and \cite{pinheiro2000mixed}, is
\begin{align*}
  \begin{gathered}
    \texttt{conc}_{ij} = \beta_{0j} + \beta_{1j}\texttt{age}_{ij} + \epsilon_{ij} \\
    \begin{pmatrix}
      \beta_{0j} \\
      \beta_{1j} \\
    \end{pmatrix}
    \sim \N \left(
      \begin{pmatrix}
        \beta_{0} \\
        \beta_{1} \\
      \end{pmatrix},
      \begin{pmatrix}
        \sigma_{0}^2 & \sigma_{01} \\
        \sigma_{01}  & \sigma_1^2 \\
      \end{pmatrix}
    \right) \\
    \epsilon_{ij} \sim \N(0, \sigma^2) \\
  \end{gathered}
\end{align*}
for $i=1,\dots,n_j$ and the index $j$ representing the 10 \code{Lots}.
Fitting this model using \code{lmer}, we can test for the significance of the fixed effect $\beta_0$, for which we find that it is not ($p$-value = 0.616), and arrive at the same conclusion as in the I-prior model.
However, we notice that the package reports a perfect negative correlation between the random effects, $\sigma_{01}$.
This indicates a potential numerical issue when fitting the model---a value of exactly $-1$, $0$ or $1$ is typically imposed by the package to force through estimation in the event of non-positive definite covariance matrices arising.
We can inspect the eigenvalues of the covariance matrix for the random effects to check that they are indeed non-positive definite.

\begin{Schunk}
\begin{Sinput}
R> (mod.lmer <- lmer(conc ~ age + (age | Lot), IGF))
\end{Sinput}
\begin{Soutput}
boundary (singular) fit: see ?isSingular
\end{Soutput}
\begin{Soutput}
Linear mixed model fit by REML ['lmerMod']
Formula: conc ~ age + (age | Lot)
   Data: IGF
REML criterion at convergence: 594.3662
Random effects:
 Groups   Name        Std.Dev. Corr 
 Lot      (Intercept) 0.082472      
          age         0.008093 -1.00
 Residual             0.820624      
Number of obs: 237, groups:  Lot, 10
Fixed Effects:
(Intercept)          age  
   5.374980    -0.002535  
convergence code 0; 1 optimizer warnings; 0 lme4 warnings 
\end{Soutput}
\begin{Sinput}
R> eigen(VarCorr(mod.lmer)$Lot)
\end{Sinput}
\begin{Soutput}
eigen() decomposition
$values
[1] 0.006867188 0.000000000

$vectors
            [,1]        [,2]
[1,] -0.99521922 -0.09766625
[2,]  0.09766625 -0.99521922
\end{Soutput}
\end{Schunk}

\begin{Schunk}
\begin{figure}[t]

{\centering \includegraphics[width=\maxwidth]{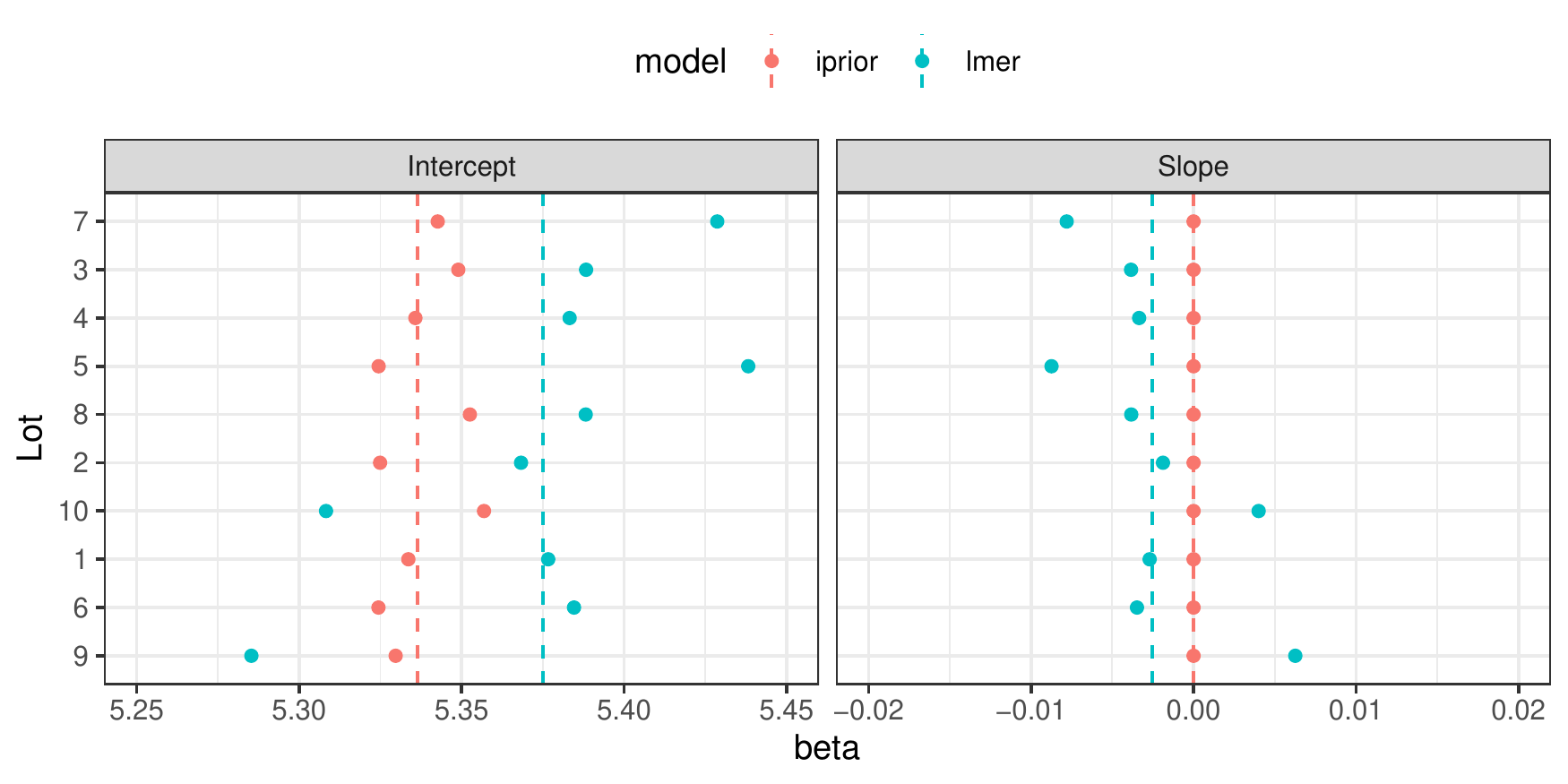} 

}

\caption[A comparison of the estimates for random intercepts and slopes (denoted as points) using the I-prior model and the standard random effects model]{A comparison of the estimates for random intercepts and slopes (denoted as points) using the I-prior model and the standard random effects model. The dashed vertical lines indicate the fixed effect values.}\label{fig:IGF.plot.beta}
\end{figure}
\end{Schunk}

\begin{table}[b!]
\centering
\begin{tabular}{lrr}
\toprule
Parameter     & \texttt{iprior} & \texttt{lmer} \\
\midrule
$\sigma_0$    & 0.012 & 0.082 \\
$\sigma_1$    & 0.000 & 0.008 \\
$\rho_{01}$   & 0.690& -1.000 \\
\bottomrule
\end{tabular}
\caption{A comparison of the estimates for the covariance matrix of the random effects using the I-prior model and the standard random effects model.}
\label{tab:igf}
\end{table}

Degenerate covariance matrices often occur in models with a large number of random coefficients.
These are typically solved by setting restrictions which then avoids overparameterising the model.
One advantage of the I-prior method for varying intercept/slopes model is that the positive-definiteness is automatically taken care of.
Furthermore, I-prior models typically require less number of parameters to fit a similar varying intercept/slopes model -- in the above example, the I-prior model estimated only three parameters, while the standard random effects model estimated a total of six parameters.

It is also possible to ``recover'' the estimates of the standard random effects model from the I-prior model, albeit in a slighly manual fashion.
Denote by $f^j$ the individual linear regression lines for each of the $j=1,\dots,10$ \code{Lots}.
Then, each of these $f^j$ has a slope and intercept for which we can estimate from the fitted values $\hat f^j(x_{ij})$, $i=1,\dots,n_j$.
This would give us the estimate of the posterior mean of the random intercepts and slopes; these would typically be obtained using empirical-Bayes methods in the case of the standard random effects model.

Furthermore, $\sigma_0^2$ and $\sigma_1^2$ gives a measure of variability of the intercepts and slopes of the different groups, and this can be calculated from the estimates of the random intercepts and slopes.
In the same spirit, $\rho_{01} = \sigma_{01} / (\sigma_0 \sigma_1)$, which is the correlation between the random intercept and slope, can be similarly calculated.
Finally, the fixed effects can be estimated from the intercept and slope of the best fit line running through the I-prior estimated \code{conc} values.
The intuition for this is that the fixed effects are essentially the ordinary least squares (OLS) of a linear model if the groupings are disregarded.
Figure \ref{fig:IGF.plot.beta} illustrates the differences in the estimates for the random coefficients, while Table \ref{tab:igf} illustrates the differences in the estimates for the covariance matrix.
Minor differences do exist, with the most noticeable one being that the slopes in the I-prior model are categorically estimated as zero, and the sign of the correlation $\rho_{01}$ being opposite in both models.
Even so, the conclusions from both models are similar.

\subsection{Longitudinal data analysis}
\label{sec:cows}

We consider a balanced longitudinal data set consisting of weights in kilograms of 60 cows, 30 of which were randomly assigned to treatment group A, and the remaining 30 to treatment group B.
The animals were weighed 11 times over a 133-day period; the first 10 measurements for each animal were made at two-week intervals and the last measurement was made one week later.
This experiment was reported by \cite{kenward1987method}, and the data set is included as part of the package \pkg{jmcm} \citep{jmcm} in \proglang{R}.
The variable names have been renamed for convenience.

\begin{Schunk}
\begin{Sinput}
R> data(cattle, package = "jmcm")
R> names(cattle) <- c("id", "time", "group", "weight")
R> cattle$id <- as.factor(cattle$id)  # convert to factors
R> str(cattle)
\end{Sinput}
\begin{Soutput}
'data.frame':	660 obs. of  4 variables:
 $ id    : Factor w/ 60 levels "1","2","3","4",..: 1 1 1 1 1 1 1 1 1 1 ...
 $ time  : num  0 14 28 42 56 70 84 98 112 126 ...
 $ group : Factor w/ 2 levels "A","B": 1 1 1 1 1 1 1 1 1 1 ...
 $ weight: int  233 224 245 258 271 287 287 287 290 293 ...
\end{Soutput}
\end{Schunk}

The response variable of interest are the \code{weight} growth curves, and the aim is to investigate whether a treatment effect is present.
The usual approach to analyse a longitudinal data set such as this one is to assume that the observed growth curves are realizations of a Gaussian process.
For example, \cite{kenward1987method} assumed a so-called ante-dependence structure of order $k$, which assumes an observation depends on the previous $k$ observations, but given these, is independent of any preceeding observations.

Using the I-prior, it is not necessary to assume the growth curves were drawn randomly.
Instead, it suffices to assume that they lie in an appropriate function class.
For this example, we assume that the function class is the fBm RKHS, i.e., we assume a smooth effect of time on weight.
The growth curves form a multidimensional (or functional) response equivalent to a ``wide'' format of representing repeated measures data. In our analysis using the \pkg{iprior} package, we used the ``long'' format and thus our (unidimensional) sample size $n$ is equal to $60$ cows $\times$ $11$ repeated measurements.
We also have two covariates potentially influencing growth, namely the cow subject \code{id} and also treatment \code{group}. The regression model can then be thought of as
\begin{align*}
  \begin{gathered}
    \text{\code{weight}} = \alpha + f(\text{\code{id}}, \, \text{\code{group}}, \, \text{\code{time}}) + \epsilon \\
    \epsilon \sim \N(0, \psi^{-1}).
  \end{gathered}
\end{align*}
\begin{table}[t!]
\centering
\begin{tabular}{lp{6cm}l}
\toprule
Model & Explanation & Formula (\verb@weight ~ ...@) \\
\midrule
1     & Growth does not vary with treatment nor among cows
&\verb@time@ \\
2     & Growth varies among cows only
&\verb@id * time@ \\
3     & Growth varies with treatment only
&\verb@group * time@ \\
4     & Growth varies with treatment and among cows
&\verb@id * time + group * time@ \\
5     & Growth varies with treatment and among cows, with an interaction effect between treatment and cows
&\verb@id * group * time@ \\
\bottomrule
\end{tabular}
\caption{A brief description of the five models fitted using I-priors.}
\label{tab:cowmodel}
\end{table}

\vspace{-1em}
We assume iid errors, and in addition to a smooth effect of \code{time}, we further assume a nominal effect of both cow \code{id} and treatment \code{group} using the Pearson RKHS.
In the \pkg{iprior} package, factor type objects are treated with the Pearson kernel automatically, and the only \code{model} option we need to specify is the \code{kernel = "fbm"} option for the \code{time} variable.
We have opted not to estimate the Hurst coefficient in the interest of computational time, and instead left it at the default value of 0.5.
Table \ref{tab:cowmodel} explains the five models we have fitted.

The simplest model fitted was one in which the growth curves do not depend on the treatment effect or individual cows.
We then added treatment effect and the cow \code{id} as covariates, separately first and then together at once.
We also assumed that both of these covariates are time-varying, and hence added also the interaction between these covariates and the \code{time} variable.
The final model was one in which an interaction between treatment effect and individual cows was assumed, which varied over time.

All models were fitted using the \code{mixed} estimation method.
Compared to the EM algorithm alone, we found that the combination of direct optimisation with the EM algorithm in the \code{mixed} routine fits the model about six times faster for this data set due to slow convergence of EM algorithm.
Here is the code and output for fitting the first model:

\begin{Schunk}
\begin{Sinput}
R> # Model 1: weight ~ f(time)
R> (mod1 <- iprior(weight ~ time, cattle, kernel = "fbm", method = "mixed"))
\end{Sinput}
\begin{Soutput}
Running 5 initial EM iterations
======================================================================
Now switching to direct optimisation
final  value 1394.615060 
converged
\end{Soutput}
\begin{Soutput}
Log-likelihood value: -2789.231 

 lambda     psi 
0.83658 0.00375 
\end{Soutput}
\end{Schunk}

\newcolumntype{R}[1]{>{\raggedleft\arraybackslash}p{#1}}
\begin{table}[t!]
\centering
\begin{tabular}{rp{4.9cm}R{2.3cm}R{1.9cm}R{2.2cm}}
\toprule
{\small Model}
& {\small{Formula \newline (}\verb@weight ~ ...@{)}}
& {\small{Log-likelihood}}
& {\small{Error S.D.}}
& {\small{Number of parameters}}  \\
\midrule
1 & \code{time}
& -2789.23
& 16.33
& 1 \\
2 & \code{id * time}
& -2789.20
& 16.32
& 2 \\
3 & \code{group * time}
& -2295.16
& 3.68
& 2 \\
4 & \code{id * time + group * time}
& -2270.85
& 3.39
& 3 \\
5 & \code{id * group * time}
& -2249.25
& 3.91
& 3 \\
\bottomrule
\end{tabular}
\caption{Summary of the five I-prior models fitted to the cow data set.}
\label{tab:cowresults}
\end{table}

\begin{Schunk}
\begin{figure}[h]

{\centering \includegraphics[width=\maxwidth]{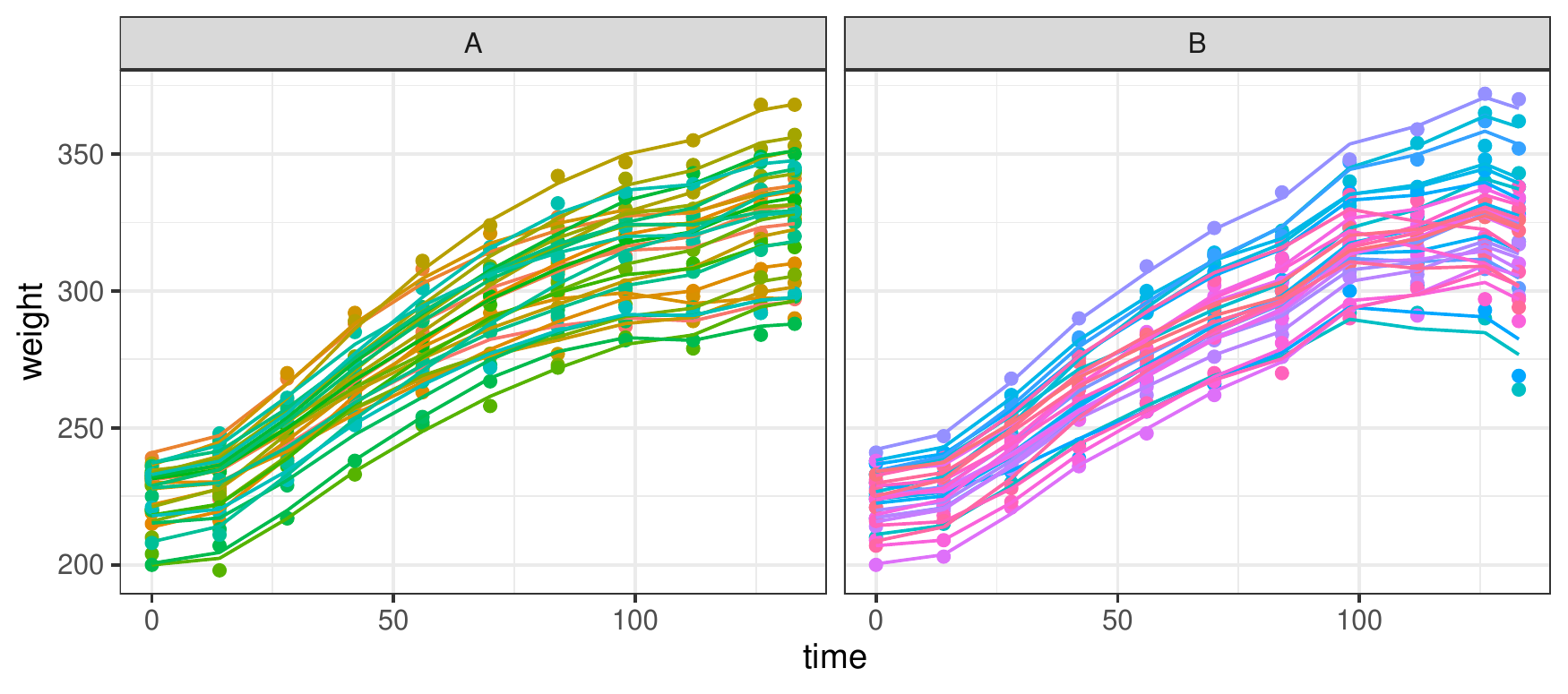} 

}

\caption[A plot of the I-prior fitted regression curves from Model 5]{A plot of the I-prior fitted regression curves from Model 5. In this model, growth curves differ among cows and by treatment effect (with an interaction between cows and treatment effect), thus producing these 60 individual lines, one for each cow, split between their respective treatment groups (A or B).}\label{fig:cows.plot}
\end{figure}
\end{Schunk}

The results of the model fit are summarised in Table \ref{tab:cowresults}. We can test for a treatment effect by testing Model 4 against the alternative that Model 2 is true.
The log-likelihood ratio test statistic is
$D = -2(-2789.20 - (-2270.85)) = 1036.70$ which has an asymptotic chi-squared distribution with $3 - 2 = 1$ degree of freedom.
The $p$-value for this likelihood ratio test is less than $10^{-6}$, so we conclude that Model 4 is significantly better than Model 2.

We can next investigate whether the treatment effect differs among cows by comparing Model 5 against Model 4.
As these models have the same number of parameters, we can simply choose the one with the higher likelihood, which is Model 5.
We conclude that treatment does indeed have an effect on growth, and that the treatment effect differs among cows.
A plot of the fitted regression curves onto the cow data set is shown in Figure \ref{fig:cows.plot}.

\subsection{Regression with a functional covariate}

We illustrate the prediction of a real valued response with a functional covariate using a widely analysed data set for quality control in the food industry.
The data\footnotemark\ contain samples of spectrometric curve of absorbances of 215 pieces of finely chopped meat, along with their water, fat and protein content.
These data are recorded on a Tecator Infratec Food and Feed Analyzer working in the wavelength range 850--1050 nm by the Near Infrared Transmission (NIT) principle.
Absorption data has not been measured continuously, but instead 100 distinct wavelengths were obtained. Figure \ref{fig:tecator.data} shows a sample of 10 such spectrometric curves.

\footnotetext{
Obtained from Tecator (see \url{http://lib.stat.cmu.edu/datasets/tecator} for details).
We used the version made available in the dataframe \code{tecator} from the \proglang{R} package \pkg{caret} \citep{caret}.
}

\begin{Schunk}
\begin{Soutput}
Warning: `as.tibble()` is deprecated, use `as_tibble()` (but mind the new semantics).
This warning is displayed once per session.
\end{Soutput}
\begin{figure}

{\centering \includegraphics[width=12cm]{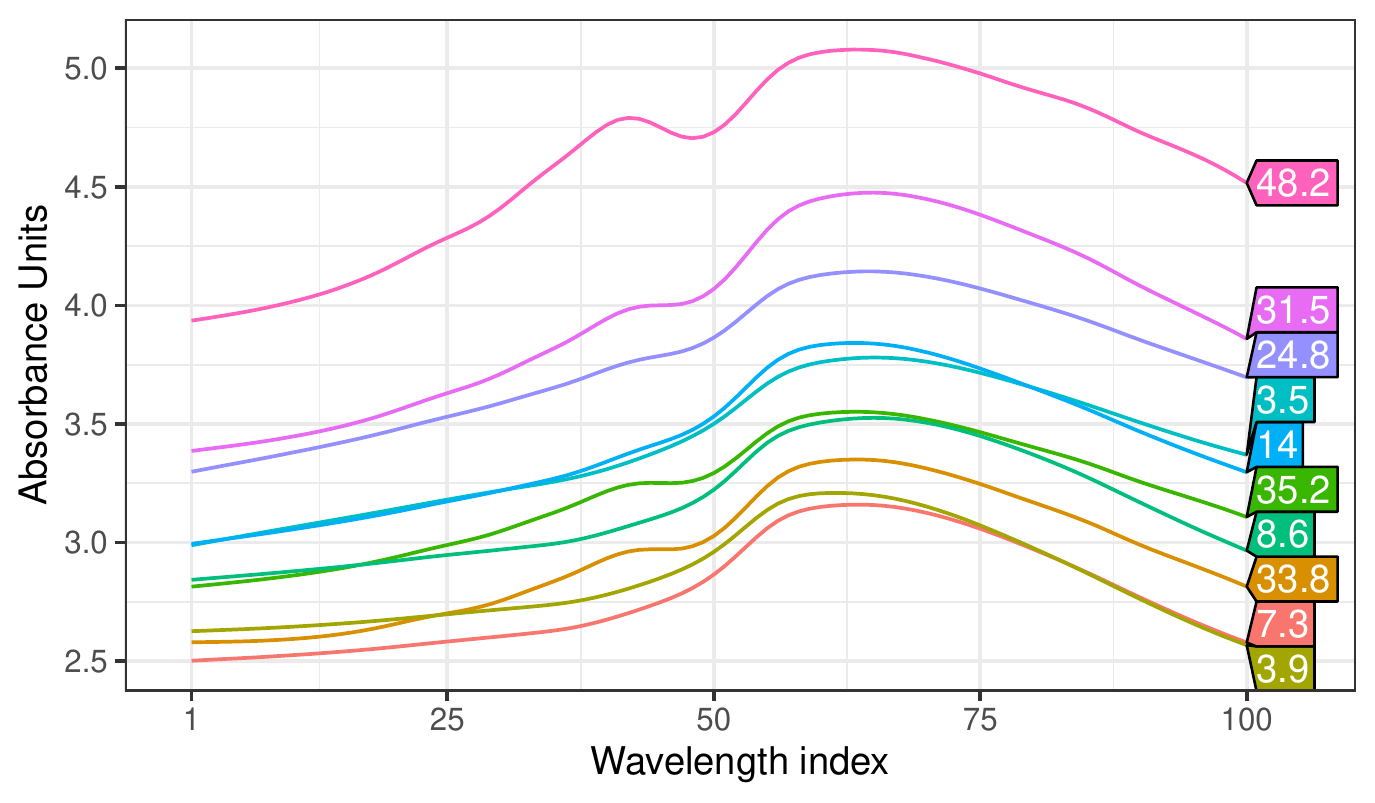} 

}

\caption[Sample of spectrometric curves used to predict fat content of meat]{Sample of spectrometric curves used to predict fat content of meat. For each meat sample the data consists of a 100 channel spectrum of absorbances and the contents of moisture, fat (numbers shown in boxes) and protein measured in percent. The absorbance is $-\log 10$ of the transmittance measured by the spectrometer. The three contents, measured in percent, are determined by analytic chemistry.}\label{fig:tecator.data}
\end{figure}
\end{Schunk}

For our analyses and many others' in the literature, the first 172 observations in the data set are used as a training sample for model fitting, and the remaining 43 observations as a test sample to evaluate the predictive performance of the fitted model.
The focus here is to use the \pkg{iprior} package to fit several I-prior models to the Tecator data set, and calculate out-of-sample predictive error rates.
We compare the predictive performance of I-prior models against Gaussian process regression and the many other different methods applied on this data set.
These methods include neural networks \citep{thodberg1996review}, kernel smoothing \citep{ferraty2006nonparametric}, single and multiple index functional regression models \citep{chen2011single}, sliced inverse regression (SIR) and sliced average variance estimation (SAVE), multivariate adaptive regression splines (MARS), partial least squares (PLS), and functional additive model with and without component selection (FAM \& CSEFAM).
An analysis of this data set using the SIR and SAVE methods were conducted by  \cite{lian2014series}, while the MARS, PLS and (CSE)FAM methods were studied by \cite{zhu2014structured}.
Table \ref{tab:tecator} tabulates the results of all of these methods from the various references.

Assuming a regression model as in \eqref{eq:linmod}, we would like to model the \code{fat} content $y_i$ using the spectral curves $x_i$.
Let $x_i(t)$ denote the absorbance for wavelength $t = 1,\dots,100$.
From Figure \ref{fig:tecator.data}, it appears that the curves are smooth enough to be differentiable, and therefore it is reasonable to assume that they lie in the Sobolev-Hilbert space as discussed in Section \ref{sec:sobolevhilbert}.
We take first differences of the 100-dimensional matrix, which leaves us with the 99-dimensional covariate saved in the object named \code{absorp}.
The \code{fat} and \code{absorp} data have been split into \code{*.train} and \code{*.test} samples, as mentioned earlier.
Our first modelling attempt is to fit a linear effect by regressing the responses \code{fat.train} against a single high-dimensional covariate \code{absorp.train} using the linear RKHS and the direct optimisation method.

\begin{Schunk}
\begin{Sinput}
R> # Model 1: Canonical RKHS (linear)
R> (mod1 <- iprior(y = fat.train, absorp.train))
\end{Sinput}
\begin{Soutput}
iter   10 value 222.653144
final  value 222.642108 
converged
\end{Soutput}
\begin{Soutput}
Log-likelihood value: -445.2844 

    lambda        psi 
4576.86595    0.11576 
\end{Soutput}
\end{Schunk}

Our second and third model uses polynomial RKHSs of degrees two and three, which allows us to model quadratic and cubic terms of the spectral curves respectively.
We also opted to estimate a suitable offset parameter, and this is called to \code{iprior()} with the option \code{est.offset = TRUE}.
Each of the two models has a single scale parameter, an offset parameter, and an error precision to be estimated.
The direct optimisation method has been used, and while both models converged regularly, it was noticed that there were multiple local optima that hindered the estimation (output omitted).

\begin{Schunk}
\begin{Sinput}
R> # Model 2: Polynomial RKHS (quadratic)
R> mod2 <- iprior(y = fat.train, absorp.train, kernel = "poly2",
+                 est.offset = TRUE)
R> # Model 3: Polynomial RKHS (cubic)
R> mod3 <- iprior(y = fat.train, absorp.train, kernel = "poly3",
+                 est.offset = TRUE)
\end{Sinput}
\end{Schunk}

Next, we attempt to fit a smooth dependence of fat content on the spectrometric curves using the fBm RKHS.
By default, the Hurst coefficient for the fBm RKHS is set to be 0.5.
However, with the option \code{est.hurst = TRUE}, the Hurst coefficient is included in the estimation procedure.
We fit models with both a fixed value for Hurst (at 0.5) and an estimated value for Hurst.
For both of these models, we encountered numerical issues when using the direct optimisation method.
The L-BFGS algorithm kept on pulling the hyperparameter towards extremely high values, which in turn made the log-likelihood value greater than the machine's largest normalised floating-point number (\code{.Machine$double.xmax = 1.797693e+308}).
Investigating further, it seems that estimates at these large values give poor training and test error rates, though likelihood values here are high (local optima).
To get around this issue, we used the EM algorithm to estimate the fixed Hurst model, and the \code{mixed} method for the estimated Hurst model.
For both models, the \code{stop.crit} was relaxed and set to \code{1e-3} for quicker convergence, though this did not affect the predictive abilities compared to a more stringent \code{stop.crit}.

\begin{Schunk}
\begin{Sinput}
R> # Model 4: fBm RKHS (default Hurst = 0.5)
R> (mod4 <- iprior(y = fat.train, absorp.train, kernel = "fbm",
+                  method = "em", control = list(stop.crit = 1e-3)))
\end{Sinput}
\begin{Soutput}
==============================================
Converged after 65 iterations.
\end{Soutput}
\begin{Soutput}
Log-likelihood value: -204.4592 

    lambda        psi 
   3.24112 1869.32897 
\end{Soutput}
\end{Schunk}
\begin{Schunk}
\begin{Sinput}
R> # Model 5: fBm RKHS (estimate Hurst)
R> (mod5 <- iprior(fat.train, absorp.train, kernel = "fbm", method = "mixed",
+                  est.hurst = TRUE, control = list(stop.crit = 1e-3)))
\end{Sinput}
\begin{Soutput}
Running 5 initial EM iterations
======================================================================
Now switching to direct optimisation
iter   10 value 115.648462
final  value 115.645800 
converged
\end{Soutput}
\begin{Soutput}
Log-likelihood value: -231.2923 

   lambda     hurst       psi 
204.97184   0.70382   9.96498 
\end{Soutput}
\end{Schunk}

Finally, we fit an I-prior model using the SE RKHS with lengthscale estimated.
Here we illustrate the use of the \code{restarts} option, in which the model is fitted repeatedly from different starting points.
In this case, eight random initial parameter values were used and these jobs were parallelised across the eight available cores of the machine.
The additional \code{par.maxit} option in the \code{control} list is an option for the maximum number of iterations that each parallel job should do.
We have set it to 100, which is the same number for \code{maxit}, but if \code{par.maxit} is less than \code{maxit}, the estimation procedure continues from the model with the best likelihood value.
We see that starting from eight different initial values, direct optimisation leads to (at least) two log-likelihood optima sites, $-231.5$ and $-680.5$.

\begin{Schunk}
\begin{Sinput}
R> # Model 6: SE kernel
R> (mod6 <- iprior(fat.train, absorp.train, est.lengthscale = TRUE,
+                  kernel = "se", control = list(restarts = TRUE,
+                                                par.maxit = 100)))
\end{Sinput}
\begin{Soutput}
Performing 8 random restarts on 8 cores
======================================================================
Log-likelihood from random starts:
    Run 1     Run 2     Run 3     Run 4     Run 5     Run 6     Run 7 
-680.4637 -680.4637 -231.5440 -231.5440 -680.4637 -680.4637 -231.5440 
    Run 8 
-231.5440 
Continuing on Run 8 
final  value 115.771932 
converged
\end{Soutput}
\begin{Soutput}
Log-likelihood value: -231.544 

     lambda lengthscale         psi 
   96.11378     0.09269     6.15424 
\end{Soutput}
\end{Schunk}

\newcolumntype{R}[1]{>{\raggedleft\arraybackslash}p{#1}}
\begin{table}[t!]
\centering
\begin{threeparttable}
\begin{tabular}{p{7cm}rr}
\toprule
\Bot &\multicolumn{2}{c}{RMSE} \\
\cline{2-3}
\Top Model & Train & Test \\
\midrule
\emph{I-prior} \\
\hspace{0.5em} Linear
& 2.89
& 2.89 \\
\hspace{0.5em} Quadratic
& 0.72
& 0.97 \\
\hspace{0.5em} Cubic
& 0.37
& 0.58 \\
\hspace{0.5em} Smooth (fBm-0.50)
& 0.00
& 0.68 \\
\hspace{0.5em} Smooth (fBm-0.70)
& 0.19
& 0.63 \\
\hspace{0.5em} Smooth (SE-0.09)
& 0.35
& 1.85 \\
\\
\emph{Gaussian process regression} \\
\hspace{0.5em} Linear
& 0.18
& 2.36 \\
\hspace{0.5em} Smooth (SE-4.89)
& 0.19
& 2.93 \\
\\
\emph{Others} \\
\hspace{0.5em} Neural network\tnote{a}                     && 0.36 \\
\hspace{0.5em} Kernel smoothing\tnote{b}                   && 1.49 \\
\hspace{0.5em} Single/multiple indices model\tnote{c}      && 1.55 \\
\hspace{0.5em} Sliced inverse regression                   && 0.90 \\
\hspace{0.5em} Sliced average variance estimation          && 1.70 \\
\hspace{0.5em} MARS\tnote{d}                               && 0.88 \\
\hspace{0.5em} Partial least squares\tnote{d}              && 1.01 \\
\hspace{0.5em} CSEFAM\tnote{d}                             && 0.85 \\
\bottomrule
\end{tabular}
\begin{tablenotes}\footnotesize
\item [a] Neural network best results with automatic relevance determination (ARD) quoted.
\item [b] Data set used was a 160/55 training/test split.
\item [c] These are results of a leave-one-out cross-validation scheme.
\item [d] Data set used was an extended version with $n=240$, and a random 185/55 training/test split.
\end{tablenotes}
\end{threeparttable}
\caption{A summary of the root mean squared error (RMSE)of prediction for the I-prior models and various other methods in literature conducted on the Tecator data set. Values for the methods under \emph{Others} were obtained from the corresponding references cited earlier.}
\label{tab:tecator}
\end{table}

Predicted values of the test data set can be obtained using the \code{predict()} function.
An example for obtaining the first model's predicted values is shown below.
The \code{predict()} method for \code{ipriorMod} objects also return the test MSE if the vector of test data is supplied.

\begin{Schunk}
\begin{Sinput}
R> predict(mod1, newdata = list(absorp.test), y.test = fat.test)
\end{Sinput}
\begin{Soutput}
Test RMSE: 2.890353 

Predicted values:
 [1] 43.607 20.444  7.821  4.491  9.044  8.564  7.935 11.615 13.807
[10] 17.359
# ... with 33 more values
\end{Soutput}
\end{Schunk}

These results are summarised in Table \ref{tab:tecator}.
For the I-prior models, a linear effect of the functional covariate gives a training RMSE of 2.89, which is improved by both the qudratic and cubic model.
The training RMSE is improved further by assuming a smooth RKHS of functions for $f$, i.e. the fBm and SE RKHSs.
When it comes to out-of-sample test error rates, the cubic model gives the best RMSE out of the I-prior models for this particular data set, with an RMSE of 0.58.
This is followed closely by the fBm RKHS with estimated Hurst coefficient (fBm-0.70) and also the fBm RKHS with default Hurst coefficient (fBm-0.50).
The best performing I-prior model is only outclassed by the neural networks of \cite{thodberg1996review}, who also performed model selection using automatic relevance determination (ARD).
The I-prior models also give much better test RMSE than Gaussian process regression\footnote{GPR models were fit using \texttt{gausspr()} in \pkg{kernlab}.}.


\section{Summary and discussion} \label{sec:summary}

The \pkg{iprior} package provides methods to estimate and analyse I-prior regression models.
Philosophically, the I-prior approach is very different from GPR in that the former starts with a function space and defines an automatic prior over that space, while the latter starts with a prior, to be chosen from prior experience or subjectively.
From a computational point of view, the two methods differ in the specification of the covariance kernel, where the I-prior approach is particularly attractive in combination with the EM algorithm.
The use of the squared Gram matrices also means that it is important that orthogonal decompositions are used, because we would be wasting computational resources in squaring the kernel matrix naively otherwise.
Further, most GPR software opt to estimate models with fixed kernel parameters---this is a crucial difference in I-prior modelling in which most, if not all, hyperparameters are estimated for inferential purposes.
We had also not come across any software package that implements the fBm kernel.

We have identified several areas of improvement, and the first is regarding estimation speed.
The nature of I-prior models means that the estimation scales as $O(n^3N)$, with $n$ being the sample size and $N$ being the number of EM or L-BFGS iterations.
As it stands, our minimal tests suggests that the \pkg{iprior} package would struggle with data sets of sizes $n \geq 5000$, unless the Nystr\"om method is used (though this is not applicable in all cases, and the Nystr\"om method's approximation quality needs to be accounted for as well).
While every care has been taken to ensure efficiency in the code, it is clear that more work needs to be done to overcome this speed issue, either from an algorithmic standpoint or in the actual code implementation.

Furthermore, storage requirements for I-prior models is a concern for large data sets, which is the second area for improvement.
The package opts to calculate and hard store the kernel matrices so that these can simply be called by the estimation methods.
Although it is possible to be more efficient by recalculating the kernel matrices from the data, or by opting to store the $n \times 1$ pre-multiplied vectors that occur most frquently e.g. $\bH_\eta \tilde\bw$ and $\bSigma_\theta^{-1}\by$, the $O(n^2)$ storage requirement still cannot be beaten as there is still an $O(n^2)$ element that needs to be elucidated and stored (temporarily) in memory.
In our case, this is the eigendecomposition routine.

Thirdly, we would like to develop the ability to handle multidimensional responses.
This was mentioned earlier in Section \ref{sec:cows}, where we saw cow growth data represented as multidimensional vectors of weights over time.
In the current version of the package, we needed to convert this into the ``long'' data format, and thus increasing the sample size from $n$ cows to $n \times T$, i.e., $T$ time points for each cow.
Keeping the data in ``wide'' format would have been more computationally efficient, and open the possibility to support more general multidimensional response regression models.

Fourthly and lastly, the package can be extended to deal with non-iid errors, i.e., $(\epsilon_1,\dots,\epsilon_n) \sim \N_n(\bzero,\bPsi^{-1})$ and $\Psi$ has a more general symmetric form.
In particular, the ability to deal with autoregressive errors would add flexibility.


%
%
%
%


\bibliography{refs}


\newpage

\begin{appendix}

\section{I-prior models as GPR models}
\label{apx:ipriorgpr}

Consider the following regression model

\begin{align}
  \begin{gathered}
    y_i = f(x_i) + \epsilon_i \\
    \epsilon_i \iid \N(0, \psi^{-1})
  \end{gathered}
\end{align}

for $i=1,\dots,n$ with $f \in \cF$ an RKHS with kernel $h$, and an I-prior on $f$, i.e.

\[
  \bff = \big(f(x_1),\dots,f(x_n) \big)^\top \sim \N_n(\bzero, \psi\bH_\eta^2).
\]

We note that among the hyperparameters of the kernel function $\eta$, there always contains a scale parameter $\lambda$ and possibly some other hyperparameters $\nu$, such that we can write the kernel function as

\[
  h_\eta(x,x') = \lambda h_\nu(x,x').
\]

Now define the kernel

\begin{align*}
   \tilde\lambda k_\nu(x,x')
   &= \psi \sum_{i=1}^n\sum_{j=1}^n h_\eta(x,x_i) h_\eta(x',x_j) \\
  &= \psi\lambda^2 \sum_{i=1}^n\sum_{j=1}^n h_\nu(x,x_i) h_\nu(x',x_j) \\
\end{align*}

By using the parameterisation $\psi\lambda^2 \mapsto \tilde\lambda$, we can treat the above kernel as having scale parameter $\tilde\lambda$ and other hyperparameters $\nu$.
Since sums of kernels are kernels and products of kernels are kernels, the $k$ defined above is also a valid kernel.
We then have

\[
  \bff \sim \N_n(\bzero, \bK)
\]

which is the familiar Gaussian process prior with $\bK_{ij} = \tilde\lambda k_\nu(x_i,x_j)$.

\section{Remark on hyperparameters and standard errors}
\label{apx:hyperparam}

In the \pkg{iprior} package, estimation of the hyperparameters, in particular the direct method using the L-BFGS algorithm, is done without any bounds constraints on the hyperparameters.
This is achieved by using the transformations listed in Table \ref{tab:transform}.
In the package internals, the transformed parameters are always referred to as \code{theta}, while the untransformed original hyperparameters are reffered to as \code{param}.
This distinction must be noted when supplying initial values for the \code{iprior()} function, as it is in the transformed parameterisation.
A helpful function included in the package is \code{check_theta()}, which reminds the form of \code{theta}:

\begin{Schunk}
\begin{Sinput}
R> mod <- kernL(circumference ~ . ^ 2, Orange, kernel = "fbm",
+               est.hurst = TRUE)
R> check_theta(mod)
\end{Sinput}
\begin{Soutput}
theta consists of 4:
lambda[1], lambda[2], qnorm(hurst[2]), log(psi)
\end{Soutput}
\end{Schunk}

\begin{table}[t!]
\centering
\begin{tabular}{lll}
\toprule
Parameter       & Transformation & Remarks \\
\midrule
Scale           & $\lambda \mapsto \log \lambda$ & Only if single scale parameter. \\
Error precision & $\psi \mapsto \log \psi$       & \\
Hurst index     & $\gamma \mapsto \Phi^{-1}(\gamma)$ & FBm kernel. $\Phi$ is CDF of $\N(0,1)$. \\
Length scale    & $l \mapsto \log l$ & SE kernel. \\
Offset          & $c \mapsto \log c$ & Polynomial kernel. \\
\bottomrule
\end{tabular}
\caption{\label{tab:transform} Hyperparameters transformations used in the package.}
\end{table}

When using maximum likelihood, the parameters may be freely transformed without affecting the optimisation procedure.
Thus, the estimates of the hyperparameters are obtained by using the respective inverse transformations in Table \ref{tab:transform}.

The standard errors however must be transformed back using the delta method.
The univariate delta method states that if the ML estimate denoted by $\hat\theta_n$ (with a dependence on the sample size $n$) converges in distribution to $\N(\theta, \sigma^2)$, then the transformed ML estimate $g(\hat\theta_n)$ converges in distribution to $\N\big(g(\theta), \sigma^2[g'(\theta)]^2\big)$, assuming $g'(\theta)$ exists and is non-zero.
Therefore, the transformed standard errors is given by $\hat\sigma g'(\hat\theta_n)$, where $\hat\sigma$ is the standard error for $\hat\theta_n$.

\section{Remark on the scale parameters}

If one or more scale parameters are (estimated to be) negative, then the reproducing kernel for the space of functions in which the regression function lives is not positive definite anymore.
It seem arbitrary to restrict the scale parameters to the positive orthant, as the sign of of the scale parameters may be informative, especially when kernels are added and multiplied (e.g. in the varying intercept/slope model).
Note that the sign of the scale parameters itself are not identified in the model (this is easily seen when having a single scale parameter in the model since the scale is squared when it appears in the likelihood) but \emph{relative signs of the scale parameters with respect to each other} is.

The space of functions with negative scale parameters is actually called a \emph{reproducing kernel Krein space} (RKKS).
Since the building blocks for the models considered are positive definite kernels, we keep speaking about RKHSs in this paper.
As with RKHSs, the user does not require any in-depth knowledge of RKKSs in order to perform I-prior modelling.

\end{appendix}
\newpage


\end{document}